\documentclass[journal]{IEEEtran}
\usepackage{amsmath,amsfonts}
\usepackage{algorithmic}
\usepackage{array}
\usepackage{textcomp}
\usepackage{stfloats}
\usepackage{url}
\usepackage{verbatim}
\usepackage{graphicx}
\def\BibTeX{{\rm B\kern-.05em{\sc i\kern-.025em b}\kern-.08em
    T\kern-.1667em\lower.7ex\hbox{E}\kern-.125emX}}
\usepackage{balance}
\usepackage{verbatim}
\usepackage{amsfonts}
\usepackage{amssymb}
\usepackage{stfloats}
\usepackage{cite}
\usepackage{soul}
\usepackage{setspace}
\usepackage{graphicx}
\usepackage{psfrag}
\usepackage{amsmath}
\usepackage{array}
\usepackage{epstopdf}
\usepackage{authblk}
\usepackage{graphicx} 
\usepackage{amsthm} 
\usepackage{lipsum}
\usepackage{verbatim} 
\usepackage{authblk}
\usepackage{mathtools}
\usepackage{cuted}
\usepackage[lined,boxed,ruled]{algorithm2e}
\usepackage{booktabs}
\usepackage{subfigure}
\usepackage[font=small]{caption} 

\usepackage{ifthen}

\usepackage[usenames]{color}

\newtheorem{Definition}{Definition}

\newtheorem{Lemma}{Lemma}

\newtheorem{Proposition}{Proposition}
\newtheorem{Remark}{Remark}

\DeclareMathOperator*{\argmax}{argmax}
\DeclareMathOperator*{\argmin}{argmin}
\DeclareMathOperator{\diag}{diag}

\begin{document}

\title{Revisiting Outage for Edge Inference Systems}
\author{ 
{Zhanwei~Wang,~\IEEEmembership{Graduate Student Member,~IEEE},
Qunsong Zeng,
~\IEEEmembership{Member,~IEEE},
Haotian~Zheng,~\IEEEmembership{Graduate~Student~Member,~IEEE},
and Kaibin~Huang,~\IEEEmembership{Fellow,~IEEE}}%
  
\thanks{Z.~Wang, Q.~Zeng, H.~Zheng, and K. Huang are with the Department of Electrical and Computer Engineering, The University of Hong Kong, Hong Kong SAR, China (Email: \{zhanweiw, qszeng, htzheng, huangkb\}@eee.hku.hk).
Corresponding authors: K. Huang; Q. Zeng.}%
\thanks{
The work was supported in part by the
Guangdong Basic and Applied Basic Research Foundation (Grant No. 2025A1515011747), Research Grants Council of the Hong Kong Special Administrative Region, China under a fellowship award (HKU RFS2122-7S04), NSFC/RGC Collaborative Research Scheme (CRS\_HKU702/24), the Areas of Excellence scheme grant (AoE/E-601/22-R), Collaborative Research Fund (C1009-22G), and the Grants 17212423 \& 17304925, and in part by the Croucher Senior Fellowship and Shenzhen-Hong Kong-Macau Technology Research Programme (Type C) (SGDX20230821091559018).}
}

\maketitle

 \begin{abstract}

 One of the key missions of sixth-generation (6G) mobile networks is to deploy large-scale artificial intelligence (AI) models at the network edge to provide remote-inference services for edge devices. 
The resultant platform, known as edge inference, will support a wide range of Internet-of-Things applications, such as autonomous driving, industrial automation, and augmented reality. 
Given the mission-critical and time-sensitive nature of these tasks, it is essential to design edge inference systems that are both reliable and capable of meeting stringent end-to-end (E2E) latency constraints. 
Existing studies, which primarily focus on communication reliability as characterized by channel outage probability, may fail to guarantee E2E performance, specifically in terms of E2E inference accuracy and latency. 
To address this limitation, we propose a theoretical framework that introduces and mathematically characterizes the inference outage (InfOut) probability, which quantifies the likelihood that the E2E inference accuracy falls below a target threshold. 
Under an E2E latency constraint, this framework establishes a fundamental tradeoff between communication overhead (i.e., uploading more feature elements) and inference reliability as quantified by the InfOut probability. 
To find a tractable way to optimize this tradeoff, we derive accurate surrogate functions for InfOut probability by applying a Gaussian approximation to the distribution of the received discriminant gain. 
Experimental results demonstrate the superiority of the proposed design over conventional communication-centric approaches in terms of E2E inference reliability.

\end{abstract}


%
 \section{Introduction}


One mission of \emph{sixth-generation} (6G) mobile networks is the widespread deployment of pre-trained \emph{artificial intelligence} (AI) models at the network edge to support ubiquitous and real-time intelligent services \cite{zhiyan_ISEA_survey,chen2024space,ZW_spectrum,Wang2025AirBreathSensing}. This emerging paradigm, known as edge inference, will serve as a platform for deploying next-generation Internet-of-Things applications, ranging from autonomous driving to industrial automation to augmented reality  \cite{GX-CM-2020,wen2023task}. 
In such a system, features extracted from sensing data are transmitted from an edge device to an edge server for remote inference using a large-scale AI model. Given that many relevant tasks are mission-critical and time-sensitive~\cite{robot_arm,wen2023taskOTA}, it is essential to develop latency-constrained edge inference systems with guaranteed performance. A primary challenge in designing such systems is the unreliable wireless links connecting edge servers and devices, as their outage events can disrupt operations and degrade performance. To address this challenge, we propose a theoretical framework in which we introduce and mathematically characterize the new definition of \emph{inference outage} (InfOut) probability. This framework establishes a fundamental \emph{communication-computation} (C$^2$) tradeoff, which is then optimized to design novel feature transmission schemes that minimize the InfOut probability.

Since fading is a fundamental characteristic of wireless channels, ensuring reliability has been a primary concern in the design of wireless communication systems from their inception. Outage probability, defined as the likelihood of a wireless link disconnecting due to deep fades, serves as a basic metric of reliability \cite{goldsmith2005wireless}. One avenue of research in reliable communications involves mathematically characterizing link-level outage probability using abstract channel models, such as space diversity \cite{Alouni_book,alouini1999capacity,Chen_tcom2023}, and Nakagami fading channels \cite{hasna2003outage,alouini2000adaptive}. This research targets various systems and scenarios, e.g., multi-hop transmission  \cite{hasna2003outage} and inaccurate channel estimation \cite{tang1999effect}. From the perspective of reliable network design, researchers have introduced the concept of network outage probability, which generalizes outage probability to account for variations in links across a network \cite{Andrews2011}. Stochastic geometry has been adopted as a tractable tool for deriving analytical expressions for network outage probability, incorporating factors such as interference, fading, and network density \cite{Haenggi2009,Dhillon2012}. Another vein of research focuses on designing techniques to cope with fading. When \emph{channel state information at the transmitter} (CSIT) is available, outages can be minimized by adapting power, modulation, and coding to the time-varying channels \cite{goldsmith1997capacity,goldsmith1997variable,goldsmith1998adaptive,vishwanath2003adaptive}. However, these adaptive approaches require accurate channel estimation and feedback, which incur additional communication overhead and latency, as well as require more complex transmitter hardware. They may not be feasible in fast-fading scenarios. When CSIT is unavailable, communication reliability can be ensured through repeated transmissions using the basic protocol of \emph{Automatic Repeat reQuest} (ARQ).

The retransmission approach has notable drawbacks, including increased communication latency and higher channel usage. These limitations create challenges in supporting mission-critical tasks with stringent deadlines, prompting the development of new techniques in the 4G and 5G eras. In particular, the breakthroughs in \emph{multiple-input multiple-output} (MIMO) communications have enabled the leveraging of space diversity to mitigate channel fading \cite{tse2005fundamentals}. Researchers have explored the fundamental tradeoff between reducing outage probability through spatial diversity and increasing transmission rates via spatial multiplexing, a relationship known as the diversity-multiplexing tradeoff \cite{zheng2003diversity}. The demand for 5G systems to support \emph{ultra-reliable and low-latency communication} (URLLC) has led to the adoption of \emph{short packet transmission} (SPT). However, the inherent conflict between achieving URLLC and maintaining high data rates means that SPT is typically suited only for low-rate, mission-critical tasks, such as transmitting control commands and basic sensing data, including humidity, temperature, and pollution levels \cite{Nallanathan-TWC-2020, zhao2023joint}. 
In this context, outages, measured by packet decoding error probability, are addressed by developing advanced SPT techniques, such as non-coherent transmission \cite{Liva-TCOM-2019}, optimal framework structures \cite{Petar-TCOM-2017}, power control \cite{Quek-TCOM-2018}, and wireless power transfer  \cite{Schmeink-JSAC-2018}. Despite these advancements, approaches designed for low-rate tasks face difficulties in ensuring the reliability of 6G edge inference systems that require data-intensive communication, such as the transmission of high-dimensional features.

In 6G, edge inference systems are designed to provide a platform for delivering remote inference services to support AI-enabled mobile applications such as sensing, autonomous driving, and robotic control~\cite{lin2023pushing,Yang2026BatchSizeFL}. As edge inference systems represent the natural convergence of AI and communication, evaluating their reliable performance requires the generalization of traditional channel outage probability to the likelihood of failing to achieve a target \emph{end-to-end} (E2E) inference accuracy, termed InfOut probability.
A mathematical study of this performance metric has not been extensively explored in the literature, as existing work has primarily focused on developing goal-oriented techniques aimed at improving the E2E performance of edge inference systems in the presence of channel distortion, as described shortly. One popular architecture for these systems, known as split inference, balances the computational load between devices and servers by flexibly dividing a pre-trained AI model into a low-complexity device sub-model for data-feature extraction and a server-side sub-model for remote inference \cite{ZJ-CoM-2020}. Existing split-inference techniques can optimize the accuracy-latency tradeoff \cite{Zhiyan-JASC-2023}, support scalable over-the-air data aggregation \cite{Zhiyan-AirPooling}, and employ progressive feature transmission to ensure high reliability \cite{RN291}. Additionally, for latency-sensitive applications, ultra-low-latency edge inference systems have been developed based on short-packet feature transmission \cite{ZW2024ultra-LoLa} and leveraging the robustness of AI models to cope with channel distortion \cite{zeng2024knowledge}. Another popular architecture for edge inference systems is joint source and channel coding (JSCC), which exploits the auto-encoder architecture to jointly train models for inference and channel coding to overcome channel noise, thereby achieving high E2E inference accuracy \cite{Task_oriented_6G_2023}. A range of relevant research has been conducted, including task-specific analog-to-digital converters \cite{Task-based_ADC_2021}, deep learning based JSCC \cite{Task-oriented_NextG_2023}, and explainable JSCC utilizing semantic channel capacity bounds \cite{Task-oriented_SC_2023}. Despite extensive efforts on algorithm designs, there is a lack of in-depth mathematical studies on the reliability of edge inference systems despite its being a fundamental topic. Results from such studies can provide performance guarantees by quantifying the worst-case inference accuracy and guide new breakthroughs in reliable edge inference, which motivates the current work.

The proposed InfOut probability depends on the interplay of randomness in propagation and the performance of \emph{deep neural networks} (DNNs)~\cite{peng2024learning}. The latter is influenced by dynamic variations in device computing capacities, model parameters, data inputs, and other factors. The reliability of DNN performance, specifically, has been investigated in the field of computer science and is typically assessed using Monte Carlo sampling \cite{zhang2019robustness}. Based on this approach, the worst-case inference performance, measured by the \emph{$k$-th percentile performance} (KPP), is quantified and subsequently enhanced through model training \cite{RN366}. On the other hand, the reliability of DNN models under attacks has been studied by evaluating the proportion of adversarial examples that successfully induce incorrect model predictions \cite{carlini2017towards}. 
The existing studies assume a stand-alone computing process within a device or server, where wireless propagation is hence irrelevant. 
In contrast, in this work, we consider latency-constrained edge inference systems over wireless links, where the reliability issue is exacerbated by additional factors such as fading and the imposition of E2E latency constraints on resource-constrained devices. Targeting such a system performing a remote classification task, the InfOut probability is defined as the probability that the E2E inference accuracy falls below a predefined threshold. Then this work presents a theoretical framework to characterize this probability. The key contributions and findings are summarized as follows. 

 \begin{itemize}
    \item \textbf{Analysis of Inference Outage Probability:}
    For analytical tractability, we assume linear classification with feature vectors extracted by the device following a \emph{Gaussian mixture model} (GMM) \cite{Yang-CVPR-2014,figueroa2019semi}. The derived results are subsequently extended and validated to design outage-minimization schemes for the more complex case of \emph{convolutional neural network} (CNN) classification with a general feature distribution. A key step in the analysis is to upper bound the InfOut probability by the probability that the \emph{discriminant gain} (DG) of channel-distorted features, which are received at the server, falls below a threshold. 
    Based on this bound, 
    we derive a tractable surrogate function for characterizing the InfOut probability by showing that the received DG for a high-dimensional feature space can be suitably modeled as a Gaussian random variable according to the Lindeberg-Feller central limit theorem. The ensuing analysis of the said bound reveals a C$^2$ tradeoff. Specifically, consider two parameters: the number of input samples for a single inference operation and the number of uploaded features for each sample. Increasing one parameter while keeping the other fixed can incur higher E2E latency but reduce the InfOut probability, and vice versa. This finding motivates the following parametric optimization to balance this tradeoff.

 \item \textbf{Inference Outage Probability Minimization:}
Directly optimizing the C$^2$ tradeoff to minimize the InfOut probability is intractable. We address this challenge by transforming the problem into one of maximizing a continuous and differentiable surrogate of the receive DG. This surrogate is shown to be a concave function of the number of uploaded features, thereby ensuring a unique optimal solution. For the more complex case of CNN classification, we define the receive DG using the inverse Gaussian Q-function related to inference accuracy and approximate its distribution as Gaussian, a method validated with real datasets. Subsequently, we design a numerical algorithm to determine the optimal number of transmitted features per sample by learning on a training dataset and employing random masking of feature vectors.
 \item \textbf{Experiments:}
The analytical results are validated using both synthetic (e.g., GMM) and real datasets (e.g., ModelNet \cite{ModelNet-Ref}). 
The designs from optimizing the C$^2$ tradeoff closely match the optimal performance from a brute-force search and outperform conventional methods prioritizing target accuracy while neglecting the channel effects on E2E accuracy.

 \end{itemize}


The remainder of this paper is organized as follows. Section \ref{Sec:ModelandMetrics} introduces the system model and defines performance metrics. In Section \ref{sec: performance_analysis}, we present the InfOut probability analysis for the edge inference system. Outage minimization strategies for linear and CNN classification are presented in Sections \ref{sec:outage_minimization_linear} and \ref{sec:outage_minimization_CNN}, respectively. Section \ref{sec:experiments} reports experimental results, while concluding remarks are provided in Section \ref{sec:conclusion}.

%
\section{Models and Metrics}\label{Sec:ModelandMetrics}

\begin{figure}[t!]
\centering
\includegraphics[width=0.85\columnwidth]{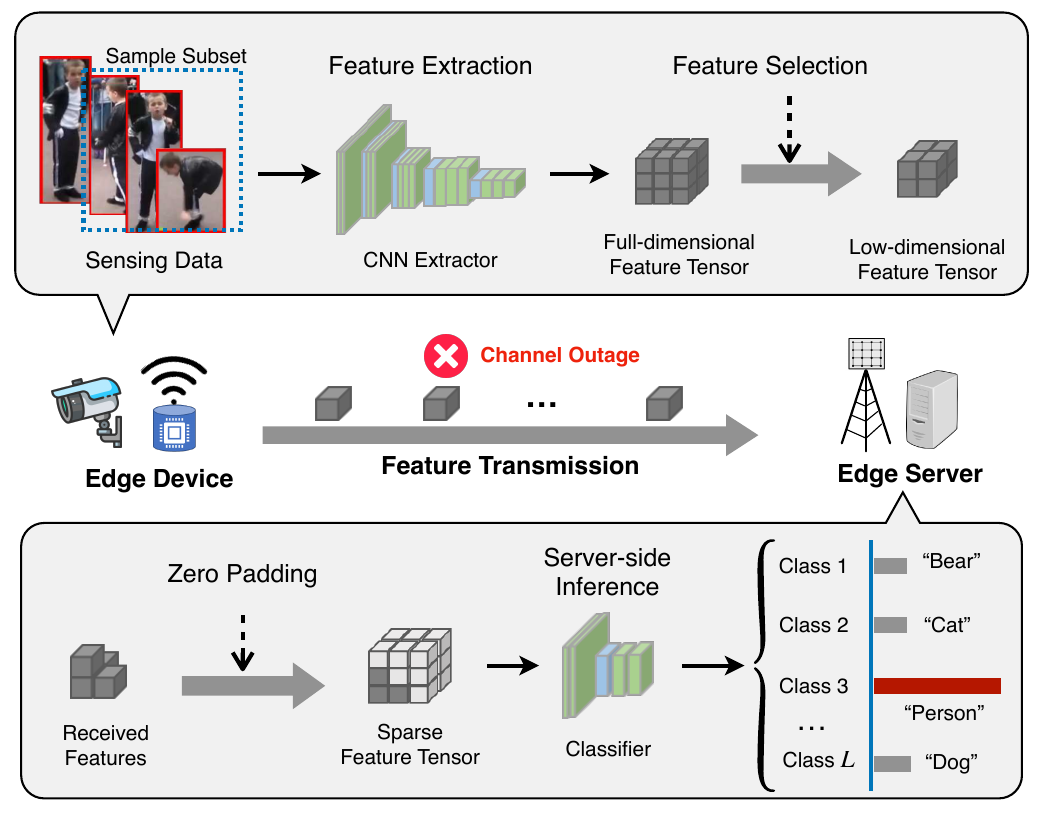}
\caption{The transceiver framework of the  edge inference system.\vspace{-2mm}}
\label{fig:framework_dig}
\end{figure}

We consider an edge inference system, as shown in Fig. \ref{fig:framework_dig}, where a device transmits feature vectors, extracted from observations, to an edge server for object classification. The associated models and performance metrics are discussed in the following subsections.

\subsection{Sensing and Computation Model}

In the considered edge inference system,
 sensing noise and potential obstructions can degrade inference performance~\cite{who2com}.
To address this issue, we consider a sensing scenario that involves fusing feature vectors extracted from $K$ observations\footnote{For instance, consider a scenario in which a device captures a series of images of a moving vehicle in an intelligent traffic monitoring system. Temporal fusion of these snapshots reduces motion blur and enhances resolution, enabling the device to transmit critical information to the server for inference~\cite{biswas2016intelligent,ZW2024ultra-LoLa}.}.
 The fused feature vector, denoted as $\overline{\mathbf{x}}$, is obtained through average pooling:
\begin{equation}
\label{eq:fea_ave}
\overline{\mathbf{x}} = \frac{1}{K}\sum_{k=1}^K \mathbf{x}_k,
\end{equation}
where $\mathbf{x}_k \in \mathbb{R}^{D}$ denotes
the feature vector extracted from the $k$-th observation.
In streaming applications, feature extraction can be pipelined across consecutive observations.
Let $N_F$ denote the number of \emph{floating-point operations }(FLOPs) required to process one observation and let $f_c$ denote the local computation speed (in FLOPs/s).
Accordingly, the per-observation feature-extraction latency is $T_{\sf fea}=N_F/f_c$.
Let $T_{\sf sen}$ denote the sensing interval (e.g., the camera frame period).
Then, under the sensing--computation pipeline, the time to obtain features from $K$ observations is
\begin{equation}
\label{eq:pipeline_process}
T_{\sf comp}=T_{\sf sen}+T_{\sf fea}+(K-1)\max\{T_{\sf sen},\,T_{\sf fea}\}.
\end{equation}
We consider a representative high–frame-rate sensing regime in which the sensing interval  $T_{\sf sen}$ is negligible (or observations are pre-buffered), so that computation for feature extraction becomes the bottleneck under a constrained power budget and limited computational speed.
With such consideration, the  computation latency in \eqref{eq:pipeline_process}  reduces to
\begin{equation}
\label{eq:computation_model}
T_{\sf comp}=\frac{K N_F}{f_c}.
\end{equation}
Note that this linear computation model is widely adopted for compute-limited edge devices with a single local processor~\cite{Mao2017MECsurvey,You2017MECO}.


\subsection{Data Distribution  Model}

While CNN based classification is designed for generic distributions, in the case of linear classification, we assume the following data distribution for tractability.
We consider feature vectors are drawn from a \emph{Gaussian mixture model} (GMM)~\cite{figueroa2019semi, RN291}.
Let $D$ and $L$ denote the dimension of the feature vector and the number of classes, respectively.
Specifically, each feature vector $\mathbf{x}_k$ is independently sampled from a Gaussian distribution with mean $\boldsymbol{\mu}_\ell \in \mathbb{R}^{D}$ and covariance matrix $\mathbf{C} \in \mathbb{R}^{D \times D}$.
The mean vector (or class centroid) varies across classes, while the covariance matrix is assumed to be identical for all classes \cite{RN291}. 
Without loss of generality, we set $\mathbf{C} = \diag(C_{1,1}, C_{2,2}, \ldots, C_{D,D})$ as a diagonal matrix, which can be obtained via \emph{principal component analysis} (PCA)~\cite{abdi2010principal}. The joint distribution of the $K$ feature vectors is given by:
\begin{equation} \label{data_dis} (\mathbf{x}_1, \ldots, \mathbf{x}_K) \sim \frac{1}{L} \sum_{\ell=1}^L \prod_{k=1}^K \mathcal{N}\left(\mathbf{x}_k \mid \boldsymbol{\mu}_\ell, \mathbf{C}\right), \end{equation}
where $\mathcal{N}\left(\mathbf{x}_k \mid \boldsymbol{\mu}_\ell, \mathbf{C}\right)$ denotes the Gaussian \emph{probability density function} (PDF) with mean $\boldsymbol{\mu}_\ell$ and covariance matrix $\mathbf{C}$.
Building on  such model, the fused feature vector, defined in \eqref{eq:fea_ave}, subjects to:
\begin{equation}
    \overline{\mathbf{x}}\sim \frac{1}{L}\sum_{\ell=1}^L \mathcal{N} (\boldsymbol{\mu}_\ell,\overline{\mathbf{C}}),
\end{equation}
where $\overline{\mathbf{C}} =\mathbf{C}/K$ with diagonal element being $\overline{C}_{d,d} = C_{d,d}/K$.
Note that the fusion of $K$ feature vectors suppresses the feature variance.

\subsection{Communication Model}

To mitigate the communication overhead from uploading high-dimensional features, we consider transmitting a subset of the fused features, denoted as $\mathcal{S}\subseteq\{1,2,\dots,D\}$, whose cardinality is $S=|\mathcal{S}|$. Each selected feature and its index are quantized into $Q_B$ and $Q_I$ bits, respectively\footnote{
Given $D$ dimensions to be indexed, the required bits per dimension are assumed to be fixed at $Q_I = \lceil \log_2(D) \rceil$.}, ensuring negligible quantization error.
Transmitting these features occupies $S$ time slots, each lasting $T_{\Delta}$ seconds.
The resulting communication latency is provided as
\begin{equation}
\label{eq:time_slot}
    T_{\sf{comm}}=T_\Delta S.
\end{equation}
For each time slot $t\in\{1,2,\dots,S\}$, 
the channel outage probability, indicating the transmission failure, is expressed as
\begin{equation}
  \label{active-prob}
  P_{\sf{out}}=\Pr(T_\Delta r_t < Q_B+Q_I).
\end{equation}
Here, $ r_t$ denotes the transmission rate, given by
\begin{equation}
    r_t=B_W \log_2\left(  1+ \frac{p|h_t|^2}{N_0B_W}\right),
\end{equation}
where $p$ is the transmit power, $B_W$ represents the system bandwidth, $N_0$ is the noise power spectrum density, and $h_t \sim \mathcal{CN}(0,\sigma^2)$ denotes the Rayleigh fading channel coefficient in the $t$-th slot. The channel is assumed i.i.d. varying over time slots but remaining constant throughout one time slot.
For convenience, we then define the \emph{activation probability} as $P_{\sf{act}}\triangleq1- P_{\sf{out}}$.
Under the assumption of i.i.d. block-fading, each feature is successfully received with probability $P_{\sf act}$.
The successfully received feature set at the edge server is denoted as $\Tilde{\mathcal{S}}\subseteq \mathcal{S} $.

\subsection{Inference Model}

We consider two classifier models based on the received feature set $\Tilde{\mathcal{S}}$.

\subsubsection{Linear Classification}
We consider a \emph{maximum likelihood} (ML) classifier for the distribution in \eqref{data_dis}, where the classification boundary between each pair of classes is a hyperplane in the feature space. 
Due to the uniform prior on the object classes, the ML classifier is equivalent to a \emph{maximum a posteriori} (MAP) classifier. The label $\hat{\ell}$ is  estimated as
\begin{equation}
\label{Maha_min_classifier}
    \begin{split}
        \hat{\ell}& =\argmax_{\ell} \log \Pr(\overline{\mathbf{x}}|\ell,\Tilde{\mathcal{S}})\\
        & = \argmin_{\ell} z_\ell(\Tilde{\mathcal{S}}),
    \end{split}  
\end{equation}
where $z_\ell(\Tilde{\mathcal{S}})$ is the squared Mahalanobis distance between the received features in $\Tilde{\mathcal{S}}$  and the centroid of class-$\ell$, given as
\begin{equation}
   z_\ell(\Tilde{\mathcal{S}})=
    \sum_{d\in \Tilde{\mathcal{S}}} \frac{(\overline{x}(d)-\mu_{\ell}(d))^2}{\overline{C}_{d,d}}.
\end{equation}
Here, $\overline{x}(d)$, $\mu_{\ell}(d)$ and $\overline{C}_{d,d}$ represent the $d$-th feature of $\overline{\mathbf{x}}$, the centroid of class-$\ell$ and the corresponding auto-covariance, respectively.
Hence, the linear classification problem reduces to finding the class label which can minimize the Mahalanobis distance.

\subsubsection{CNN Classification}
We also consider a more realistic but analytically intractable scenario where feature vectors are extracted from observations using a well-trained CNN model.
The layers of CNN model are split into a device sub-model and a server sub-model, represented as functions $f_{\sf sen}(\cdot)$ and $f_{\sf ser}(\cdot)$, respectively.
The feature vector of the $k$-th observation is constructed by passing the observation ${\mathbf{M}_k}$ of the common object through a pre-trained CNN, i.e., $\mathbf{x}_k=f_{\sf sen}(\mathbf{M}_k)$.
The feature vectors are then aggregated to $\overline{\mathbf{x}}=\frac{1}{K}\sum_{k=1}^K\mathbf{x}_k$ before feature selection and transmission.
Upon receiving the feature elements and their associated indices, the edge server reconstructs the feature map into its original dimensionality  $D$ by zero-padding any unselected and channel-lost features.
Let  $\Tilde{\mathbf{x}}_{\sf{cnn}} \in \mathbb{R}^{D}$ denote the output of this feature reconstruction.
Subsequently, the edge server feeds the sparse feature map  $\Tilde{\mathbf{x}}_{\sf{cnn}}$ into the server sub-model, i.e., $ \{c_1,\dots,c_\ell,\dots,c_L\} =f_{\sf ser}\left(\Tilde{\mathbf{x}}_{\sf{cnn}}\right)$, where $c_\ell$ represents the confidence score of the $\ell$-th class.
The CNN classifier then outputs the inferred label with the highest confidence score, i.e., $\hat{\ell}=\argmax_{\ell} c_\ell$.

\subsection{Relevant Metrics}

For linear and CNN classification, relevant metrics are characterized as follows.

\subsubsection{Inference Outage Probability}

For a classification task with  $K$ processed observations and a set of received features $\Tilde{\mathcal{S}}$, the
inference accuracy, denoted as $a(K, \Tilde{\mathcal{S}})$, is commonly defined as the probability of correctly predicting  the object label, expressed as
\begin{equation} 
\label{eq:inferece_observation_fea}
a (K, \Tilde{\mathcal{S}})=\frac{1}{L}\sum_{\ell=1}^L\Pr(\hat{\ell}=\ell \mid \ell, K, \Tilde{\mathcal{S}}).
\end{equation}
Due to the feature loss caused by block-fading channels, the set of received features $\Tilde{\mathcal{S}}$ is random, making the inference accuracy a random variable over wireless channels.
To capture the channel-induced randomness of the E2E performance, 
we assume that $a$ follows the distribution $a \sim \mathcal{D}_{\theta}(K,\mathcal{S})$, where $\mathcal{D}$ denotes the distribution of inference accuracy,  $\theta$ represents the distribution’s parameter,  $K$ is the number of processed observations, and $\mathcal{S}$ is the selected feature set at the sensor. 
Given a target inference accuracy $A_{\sf{th}}$, an inference outage occurs if the accuracy requirement is not met.
In this context, the InfOut probability, which measures the reliability of the system, is denoted as 
\begin{equation}
\label{eq:P_out}
\begin{split}
     P_{\sf{out}}^{\sf{e2e}} &=  \Pr(a  \leq A_{\sf{th}}\mid K,\mathcal{S})\\
     &= \sum_{\Tilde{\mathcal{S}} \subseteq \mathcal{S}} \mathbb{I}(a (K, \Tilde{\mathcal{S}}) \leq A_{\sf th}  ) P(\Tilde{\mathcal{S}}),
\end{split}
\end{equation}
where 
$\mathbb{I}(\cdot)$ denotes the indicator function and $P(\Tilde{\mathcal{S}})$ is the \emph{probability mass function} (PMF) of the received feature set $\Tilde{\mathcal{S}}$ at the edge server\footnote{While we focus on classification as a representative inference task, the considered metric readily extends to other inference tasks. Specifically, for regression, the InfOut probability can be defined as the probability that the \emph{mean squared error} (MSE) exceeds a target threshold; for detection, it can be defined as the probability that the \emph{intersection-over-union} (IoU) falls below a prescribed threshold.
}.




\subsubsection{On-device Feature Importance}
We consider two types of metrics to measure the on-device feature importance for linear and CNN classifications, respectively.

\begin{itemize}
    \item \emph{Discriminant Gain:}
\label{sec:dg}
For linear classification, the pairwise DG quantifies the discernibility between two classes within a subspace of the feature space.
Given the fused feature vector $\overline{\mathbf{x}}$, the DG between  class $\ell$ and $\ell'$, denoted as  $ G_{\ell,{\ell'}}$, is defined as the symmetric \emph{Kullback-Leibler} (KL) divergence~\cite{RN291}:
\begin{equation}
\begin{split}
    G_{\ell,{\ell'}} = & \sf{KL}(\mathcal{N}(\boldsymbol{\mu}_\ell,\overline{\mathbf{C}})\,||\, \mathcal{N}(\boldsymbol{\mu}_{\ell'},\overline{\mathbf{C}}))\\
    &+\sf{KL}(\mathcal{N}(\boldsymbol{\mu}_{\ell'},\overline{\mathbf{C}})\,||\,\mathcal{N}(\boldsymbol{\mu}_\ell,\overline{\mathbf{C}}))\\
     =&(\boldsymbol{\mu}_{\ell}-\boldsymbol{\mu}_{\ell'})^{\sf{T}}\overline{\mathbf{C}}^{-1}(\boldsymbol{\mu}_{\ell}-\boldsymbol{\mu}_{\ell'}) \\
     =&
 K \sum_{d=1}^D  W_d(\ell,\ell'),
\end{split}
\end{equation}
where $W_d(\ell,\ell')$ is the pair-wise DG of the $d$-th dimension, given as
\begin{equation}
\label{eq:W_d}
    W_d(\ell,\ell')=\frac{(\mu_{\ell}(d)-\mu_{\ell'}(d))^2}{C_{d,d}}.
\end{equation}

Using this metric, we quantify the importance of each feature dimension and enable the DG based feature selection scheme.
Specifically, 
the importance of the $d$-th dimension is measured by the minimum DG among all class pairs, defined as
\begin{equation}
\label{eq:min_DG}
    \hat{W}_d = \min_{\ell \neq \ell'} W_d(\ell, \ell').
\end{equation}
The min-DG rule prioritizes features that strengthen the most confusable class pair, aligning with our objective of ensuring worst-case inference performance.

\item \emph{Feature Magnitude:}
\label{sec:mag}
The DG defined
in \eqref{eq:min_DG} for a linear classifier, which underpins the associated metric of feature importance, is not applicable to a CNN model.
In this context, we adopt a magnitude based feature selection scheme, where the importance of each feature element is determined by its magnitude  \cite{feature_pruning}.
Given a target number of selected features, $S=|\mathcal{S}|$, the device selects the top-$S$ features with the largest magnitudes for transmission.
Since each selected feature is transmitted in a fixed-duration slot with a fixed power budget, the top-$S$ selection does not change the per-feature transmit energy, but only the feature dimensions that are delivered.
\end{itemize}


\subsubsection{E2E Inference Latency}
\label{subsec:e2e_latency}
We measure the task completion time by the E2E inference latency, which captures the latency accrued by the sequential inference pipeline.
Given the computation and the feature-transmission latencies in~\eqref{eq:computation_model} and~\eqref{eq:time_slot}, the E2E latency is given by
\begin{equation}
    T_{\sf e2e}=T_{\sf comp}+T_{\sf comm}.
\end{equation}
For time-sensitive applications (e.g., collision avoidance in autonomous driving), we enforce a strict latency requirement $ T_{\sf e2e}\le T_{\max}$
where $T_{\max}$ denotes the target E2E task deadline.

\section{Analysis of Inference Outage Probability}
\label{sec: performance_analysis}

This section provides a theoretical analysis of the InfOut probability for linear classification. 
The derived insights are further applied to minimizing the InfOut probability in CNN based classification, as discussed in Sec. \ref{sec:outage_minimization_CNN}.

\subsection{Tractable Surrogate of Inference Accuracy}

The computation of the InfOut probability in \eqref{eq:P_out} requires an accurate characterization of the inference accuracy distribution. For tractability, we consider the lower bound of inference accuracy provided in Lemma \ref{inference_accuracy_LB} as its surrogate.

\begin{Lemma}[\cite{ZW2024ultra-LoLa}]
\label{inference_accuracy_LB}
The inference accuracy with $K$ observations and   received feature set $\Tilde{\mathcal{S}}$, denoted as $a (K,\Tilde{\mathcal{S}})$, is lower bounded by
\begin{equation}
    \begin{split}
     a  (K, \Tilde{\mathcal{S}}) \geq a_{\sf{low}}(K,G_{\sf{R}})  \triangleq 1-(L-1)Q\left( \frac{\sqrt{K G_{\sf R}}}{2}\right),
\label{eq:inference_accuracy_LB}
  \end{split}
\end{equation}
where $G_{\sf R}$ is defined as the receive DG per observation: 
\begin{equation}
    \label{eq:rece_DG}
    G_{\sf R} =  \sum_{d\in \Tilde{\mathcal{S}}} \hat{W}_d.
\end{equation}
$\hat{W}_d$ is the minimum DG of the $d$-th feature dimension in \eqref{eq:min_DG}.
\end{Lemma}
Lemma \ref{inference_accuracy_LB} indicates that the lower bound of inference accuracy is a monotonically increasing function of the receive DG, as defined in \eqref{eq:rece_DG}.
The value of $G_{\sf R}$ increases as the number of successfully received features, denoted as $\Tilde{S}=|\Tilde{\mathcal{S}}|$, grows due to the positive DG per dimension, i.e., $\hat{W}_d\geq 0$.
However, feature loss caused by fading channels introduces randomness into $\Tilde{\mathcal{S}}$, resulting in a distribution of the receive DG and variability in inference accuracy. 
This uncertainty raises reliability concerns for edge inference systems.

By leveraging the one-to-one mapping between the lower-bounded inference accuracy and the receive DG defined in \eqref{eq:inference_accuracy_LB}, the InfOut probability in \eqref{eq:P_out} can be upper-bounded as:
\begin{equation}
\label{eq:P_out_multiclass}
    \begin{split}  
P_{\sf{out}}^{\sf{e2e}}  &=  \sum_{\Tilde{\mathcal{S}} \subseteq \mathcal{S}}  \mathbb{I}(a (K, \Tilde{\mathcal{S}})\leq A_{\sf th} ) P(\Tilde{\mathcal{S}}) \\
& =1-  \sum_{\Tilde{\mathcal{S}} \subseteq \mathcal{S}} \mathbb{I}(a (K, \Tilde{\mathcal{S}})> A_{\sf th}) P(\Tilde{\mathcal{S}}) \\
& \leq 1-  \sum_{\Tilde{\mathcal{S}} \subseteq \mathcal{S}}  \mathbb{I}(a_{\sf low}(K,G_{\sf{R}}) > A_{\sf th}) P(\Tilde{\mathcal{S}}) \\
   & =  \sum_{\Tilde{\mathcal{S}} \subseteq \mathcal{S}}   \mathbb{I}(KG_{\sf{R}}\leq  G_{\sf th}) P(\Tilde{\mathcal{S}}) \\
   &= \Pr (KG_{\sf{R}}\leq G_{\sf th} ),
   \end{split}
\end{equation}
where $G_{\sf{th}} \triangleq 4\left(Q^{-1}\left( \frac{1-A_{\sf{th}}}{L-1}  \right)\right)^2$  denotes the required DG threshold to achieve an inference accuracy of $A_{\sf{th}}$.
The result in \eqref{eq:P_out_multiclass} demonstrates that the receive DG, $G_{\sf R}$, can serve as a tractable surrogate for inference accuracy.

\subsection{Inference Outage Probability}

Without loss of generality, we assume that the DG values are arranged in decreasing order after PCA, i.e., $\hat{W}_d\geq \hat{W}_{d+1}, d=1,\cdots,D-1$.
We consider a DG based feature selection scheme that selects the top-$S$ features with the highest DG values. In such manner, the receive DG in \eqref{eq:rece_DG} can be rewritten as 
\begin{equation}
\label{eq:rece_DG_V2}
     G_{\sf R}=
    \sum_{d\in \Tilde{\mathcal{S}}} \hat{W}_d
   =\sum_{d=1}^S \hat{W}_d I_d,
\end{equation}
where $I_d$ is an indicator representing the successful transmission of the $d$-th feature dimension
over fading channels. We assume i.i.d. fading across feature slots (e.g., via time--frequency
interleaving beyond the channel coherence time and bandwidth), such that $\{I_d\}$ are i.i.d.\ Bernoulli:
\begin{equation}
    I_d=\begin{cases}
1, & \text{with probability of} \quad P_{\sf{act}},\\
0, & \text{with probability of} \quad 1-P_{\sf{act}}.
    \end{cases}
\end{equation}
Notably, $G_{\sf R}$ is the weighted sum of i.i.d Bernoulli random variables, with its mean and variance given by
\begin{equation}
\begin{split}
     \label{eq:rece_DG_mean_var}
   \mathbb{E}[G_{\sf R}]&=P_{\sf act}G_1(S), \\
  \text{Var}(G_{\sf R})&=(1-P_{\sf act})P_{\sf act}G_2(S),
\end{split}
\end{equation}
where $G_1(S)$ and $G_2(S)$ are functions of the number of selected features $S$:
\begin{equation}
\label{transmit-DG-power}
    G_1(S)=\sum_{d=1}^S \hat{W}_d, \quad G_2(S)=\sum_{d=1}^S \hat{W}^2_d.
\end{equation}
Here, we refer to $G_1(S)$ as \emph{transmit DG}, which quantifies the DG of selected features at the sensor side.
Meanwhile, $G_2(S)$ represents the sum of squared dimension-wise DGs, termed the \emph{transmit DG power}.

\begin{figure}[t!]
\centering
\subfigure[Discriminant Gain ($\hat{W}_d$)]{
\includegraphics[width=0.45\columnwidth]{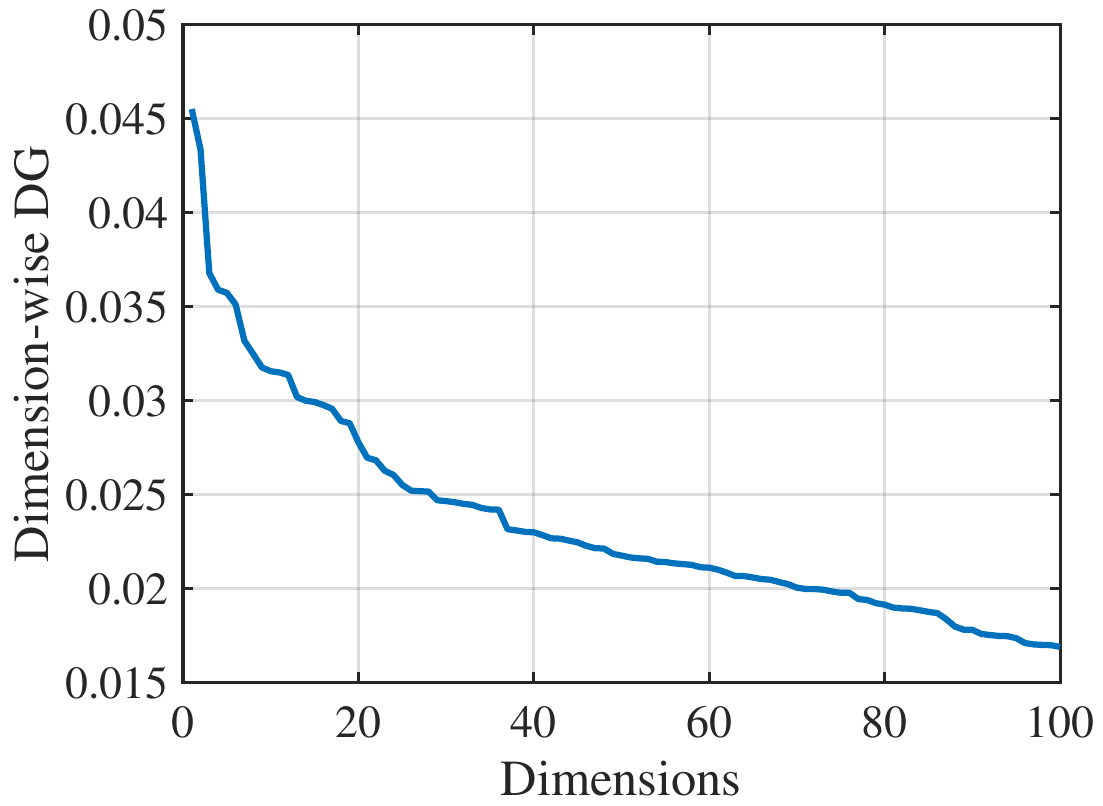}}
\subfigure[Gaussian Approximation]{
\includegraphics[width=0.42\columnwidth]{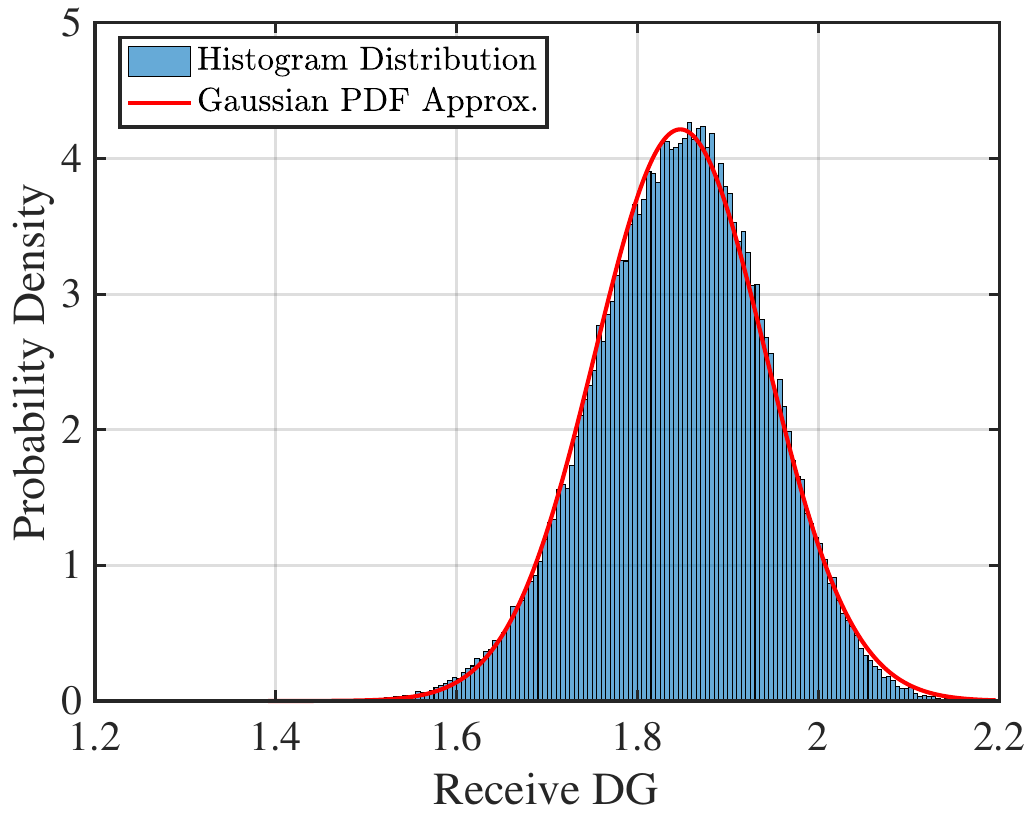}}
\caption{
Considering dimension-wise DG $\hat W_d$ sorted in descending order in subfigure (a), the PDF comparison between the real distribution and the Gaussian approximation is presented in subfigure (b). 
The number of selected features is set at $S=100$ under an activation probability of $P_{\sf{act}}=0.8$.
\vspace{-3mm}}
\label{fig:DG_approx}
\end{figure}

To characterize the distribution of $G_{\sf R}$, we show that the weighted sum of i.i.d. Bernoulli random variables in \eqref{eq:rece_DG_V2} satisfies Lindeberg's condition, as stated in Lemma \ref{Lemma:Lindeberg-con}.

\begin{Lemma}[Lindeberg's Condition \cite{feller1991introduction}]  
\label{Lemma:Lindeberg-con}  
Let \( X_d = \hat{W}_d I_d \), \( d = 1, 2, \ldots, S \) be independent random variables with mean \( \mu_{X_d} = \hat{W}_d P_{\mathsf{act}} \) and variance \( \mathrm{Var}(X_d) = P_{\mathsf{act}}(1 - P_{\mathsf{act}}) \hat{W}_d^2 \). For any \( \epsilon > 0 \), the Lindeberg condition holds:  
\begin{equation} 
\begin{split}
\label{eq:Lindeberg-con}  
&\lim_{S \to \infty} \frac{1}{\sigma_{\mathsf{G}}^2(S)} \sum_{d=1}^S \mathbb{E}\left[ \left( X_d - \mu_{X_d} \right)^2 \mathbb{I}\left( \left| X_d - \mu_{X_d} \right| > \epsilon \sigma_{\mathsf{G}}(S) \right)\right] \\
&= 0,  
\end{split}
\end{equation}  
where $\sigma_{\mathsf{G}}^2(S) =
\sum_{d=1}^S \text{Var}(X_d) =P_{\mathsf{act}}(1 - P_{\mathsf{act}}) \sum_{d=1}^S \hat{W}_d^2$
denotes the aggregate variance.  
\end{Lemma}
\noindent The proof is provided in Appendix \ref{proof:Lindeberg-con}.

Consequently, the receive DG can be approximated by a Gaussian distribution using the  Lindeberg-Feller Central Limit Theorem, as provided in Lemma \ref{DG_approx}.

\begin{Lemma}[Distribution of Receive DG]
    \label{DG_approx}
If the Lindeberg condition in Lemma \ref{Lemma:Lindeberg-con} holds, 
then for a sufficiently large number of selected features $S$ (a typical scenario in DNNs), the distribution of receive DG in \eqref{eq:rece_DG_V2} can be approximated as 
    \begin{equation}
         G_{\sf R} \rightarrow
         \mathcal{N}\left(  \mathbb{E}[G_{\sf R}], \text{Var}(G_{\sf R})\right),\quad   \text{weakly as} \quad  S\rightarrow\infty,
    \end{equation}
where $ \mathbb{E}[G_{\sf R}]$ and $\text{Var}(G_{\sf R})$ are the mean and variance of the distribution, provided in \eqref{eq:rece_DG_mean_var}.
\end{Lemma}

To illustrate the approximation, Fig. \ref{fig:DG_approx} shows the statistics of the receive DG, computed using 100 dimensions with an activation probability of $P_{\sf act}=0.8$. It can be observed that the approximation closely matches the empirical distribution.
Using this approximation, the upper bound of the InfOut probability in \eqref{eq:P_out_multiclass} can be expressed as
\begin{align}
    P_{\sf{out}}^{\sf{e2e}}  & \leq \Pr (KG_{\sf{R}}\leq G_{\sf th}) \label{eq:P_out_upperbound}\\ 
    &\approx  Q\left( \frac{ P_{\sf{act}} G_{\sf{f}}(S) -\frac{G_{\sf{th}}}{K\sqrt{G_2(S)}}
    }{\sqrt{P_{\sf{act}}(1-P_{\sf{act}})}}  \right), \label{eq:outage_prob}
\end{align}
where $Q(x)=\int_{x}^{\infty} \frac{1}{\sqrt{2\pi}} \exp{\left(-\frac{t^2}{2}\right)} d t $ denotes the  Q-function, and $G_{\sf{f}}(S) =  \frac{G_1(S)}{\sqrt{G_2(S)}}$ is a monotone-increasing function of the number of selected features $S$ (see Appendix \ref{Proof:Monotone_G_F}). 


\begin{Remark}
[Computation-communication tradeoff]
\label{remark:C2_tradeoff}
The result in \eqref{eq:outage_prob} theoretically shows that the InfOut probability can be reduced by increasing the number of selected features and/or the number of processed observations. However, this reduction comes at the cost of increased communication and/or computation latency, respectively.
Setting a strict deadline for E2E latency introduces a competition between computation and communication: allocating more time for processing more observations produces a higher quality feature vector, while fewer features can be uploaded in the reduced time available for communication, and vice versa.
Thus, a fundamental \emph{communication-computation} (C$^2$) tradeoff emerges.
\end{Remark}

\begin{figure}[t!]
\centering
\includegraphics[width=0.65\columnwidth]{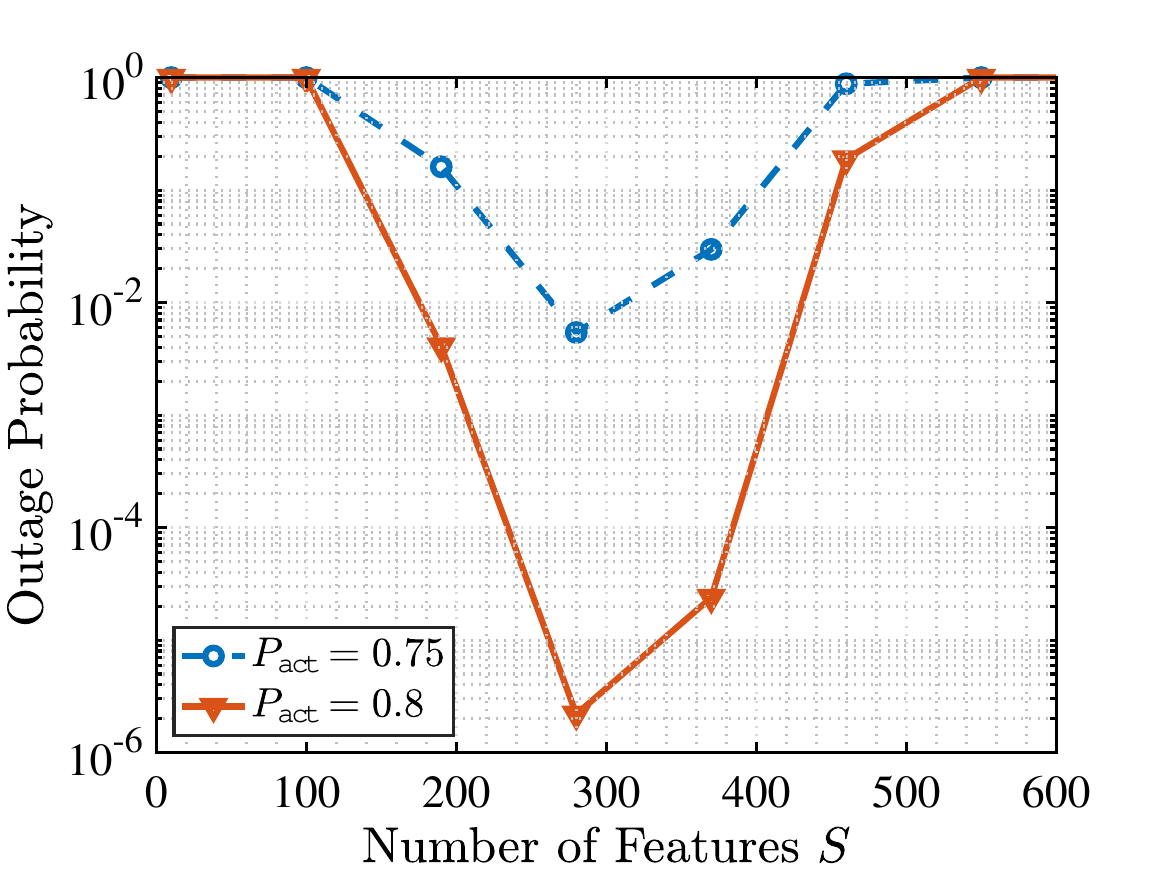}
\caption{InfOut probability under different channel outage probabilities. 
The numerical settings of a binary classifier are $A_{\sf th}=99\%$, $T_\Delta=10$~$\mu$s, $T_{\max}=10$~ms, $f_c=2.5$~GFLOPs/s, and $N_F=2.3$~MFLOPs.\vspace{-3mm}}
\label{fig:tradeoff}
\end{figure}

Fig.~\ref{fig:tradeoff} validates the C$^2$ tradeoff controlled by the number of selected features under a latency constraint of 10 ms. The InfOut probability initially decreases and subsequently increases as the number of selected features grows. Additionally, a more reliable channel (with higher activation probability $P_{\sf act}$) achieves a lower InfOut probability, highlighting the interplay between channel outage and inference outage.


\section{Minimization of Inference Outage Probability}
\label{sec:outage_minimization_linear}

In this section, we enhance the reliability of latency-constrained edge inference systems by optimizing the C$^2$ tradeoff described in Remark \ref{remark:C2_tradeoff}. The identified C$^2$ tradeoff is governed by the number of processed observations and transmitted features under E2E latency constraints.
To minimize the InfOut probability while satisfying  the latency requirement, these control variables are jointly optimized. The resulting optimization problem is formulated as
\begin{subequations}
\label{prob:outage_mini}
\begin{align}
        \min_{S,K} & \quad Q\left( \frac{ P_{\sf{act}} G_{\sf{f}}(S) -\frac{G_{\sf{th}}}{K\sqrt{G_2(S)}}
    }{\sqrt{P_{\sf{act}}(1-P_{\sf{act}})}}  \right) \label{eq:opt_obj} \\
            \mathrm{s.t.} & \quad  \frac{KN_F}{f_c} +T_\Delta S\leq T_{\max}, \label{eq:e2e_latency}\\
            & \quad S\in \{1,2,\dots, D\}, \label{eq:feature_range}\\
            & \quad  K\in \{1,2,\dots, K_{\max}\}, \label{eq:view_range}
\end{align}
\end{subequations}
where $K_{\max}$  denotes the maximum number of observations of the object.
Constraint \eqref{eq:e2e_latency} accounts for the device-side feature extraction and uplink transmission latency.
The server-side inference latency is assumed to be negligible due to sufficient computing resources at the edge server, or treated as a constant that can be absorbed into the E2E deadline.
To solve problem~\eqref{prob:outage_mini}, we approximate the objective using a surrogate function derived from the lower bounds on the transmit DG $G_1(S)$ and its associated power $G_2(S)$. Subsequently, the optimal number of selected features is determined under the DG based feature selection scheme.

\subsection{Lower Bounds on Transmit DG}
\label{sec:continous_DG}
The impact of the number of selected features on the objective of Problem \eqref{prob:outage_mini} is captured by two discrete functions, $G_1(S)$ and $G_2(S)$, where $S \in \{1, 2, \dots, D\}$. To facilitate the analysis of these functions, we derive their lower bounds by leveraging integrals of the continuous and differentiable DG function, as defined in Definition \ref{def:DG_Function}.

\begin{Definition}[Discriminant Gain Function]
\label{def:DG_Function}
The DG function is defined as  a continuous and differentiable function of dimension index $t\in [0,D]$, given as
\begin{equation}
\label{eq:DG_function}
    g\left(t\right)=
        \frac{\hat{W}_{d}-\hat{W}_{d+1}}{2}\cos(\pi (t-d+1))+\frac{\hat{W}_d+\hat{W}_{d+1}}{2}, 
\end{equation}
where  $\hat{W}_d$ denotes the DG of the $d$-th dimension in \eqref{eq:min_DG}, arranged in decreasing order such that $ \hat{W}_{d}\geq \hat{W}_{d+1}, \forall d\in \{1,2,\dots,D-1\}$. Otherwise, $\forall t \notin [0,D]$, $g(t)=0.$
\end{Definition}

\begin{figure}[t!]
\centering
\includegraphics[width=0.55\columnwidth]{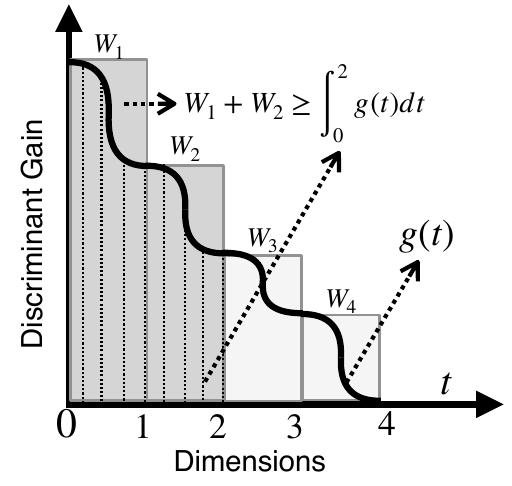}
\caption{An example of DG function with $D=4$.\vspace{-2mm}}
\label{fig:DG_approx_func}
\end{figure}

The defined DG function establishes an approximate relationship between the dimension index $d$ and the corresponding dimension-wise DG. By leveraging Definition \ref{def:DG_Function}, the functions $G_1(S)$ and $G_2(S)$ can be lower bounded through the integration of the DG function $g(t)$, denoted as $\hat{G}_1(S)$ and $\hat{G}_2(S)$, respectively, expressed as:
\begin{equation}
\label{eq:sum_DF_function}
\begin{split}
G_1(S)=\sum_{d=1}^S W_d&\geq\int_{0}^S g(t) dt\triangleq\hat{G}_1(S),\\
G_2(S)=\sum_{d=1}^S W^2_d&\geq\int_{0}^S g^2(t) dt\triangleq\hat{G}_2(S), 
 \end{split}
\end{equation}
where $S \in [1,D]$ is treated as a continuous variable, relaxing the discrete constraint on the number of selected features.
An example of the DG function $g(t)$ and its associated lower bounds in 
\eqref{eq:sum_DF_function} is illustrated in Fig. \ref{fig:DG_approx_func}.

\subsection{Optimal DG based Feature Selection}
\label{Sec:surrogate_tradeoff}

The lower bounds on transmit DG and its associated power enable the derivation of a surrogate function for the objective in Problem \eqref{prob:outage_mini}.
Since increasing either the number of selected features, $S$, or the number of processed observations, 
$K$, reduces the InfOut probability, the constraint in \eqref{eq:e2e_latency} must hold with equality. Accordingly, the maximum number of observations, denoted by $\hat{K}$, is derived from the constraint in \eqref{eq:e2e_latency} and is given by:
\begin{equation}
\label{eq:view-feature relation}
   \hat{K}=\lfloor-B_1S+B_0\rfloor,
\end{equation}
where $B_1=\frac{ f_cT_\Delta}{N_F}>0,B_0=\frac{f_cT_{\max}}{N_F}>0$.

We then relax $ \hat{K}$ by allowing it to take non-integer values.
Incorporating this relaxation and lower bounds in \eqref{eq:sum_DF_function}, we approximate the upper bound of the InfOut probability in \eqref{eq:opt_obj} as a function of the number of selected features $S$, given by
\begin{equation}
\label{eq:outage_prob1}
\begin{split}
      Q\left( \frac{ P_{\sf{act}} G_{\sf{f}}(S) -\frac{G_{\sf{th}}}{K\sqrt{G_2(S)}}
    }{\sqrt{P_{\sf{act}}(1-P_{\sf{act}})}}  \right)
    \approx Q\left( \frac{f(S)}{\sqrt{P_{\sf{act}}(1-P_{\sf{act}})}}  \right).
    \end{split}
\end{equation}
Here, $f(S)$ represents the surrogate function obtained by substituting  $G_{\sf f}=\frac{G_1(S)}{\sqrt{G_2(S)}}\approx \frac{\hat{G}_1(S)}{\sqrt{\hat{G}_2(S)}} $ and $K\approx \hat{K}$ into the expression of the numerator in Q-function, given by
\begin{equation}
\label{eq:f(x)_main}
   f(S)= \frac{P_{\sf{act}}\hat{G}_1(S)}{\sqrt{\hat{G}_2(S)}}-\frac{G_{\sf{th}}}{(B_0-B_1 S)\sqrt{\hat{G}_2(S)}}.
\end{equation}
It is obvious that InfOut probability is a monotonically decreasing function of $f(S)$. Consequently, the InfOut probability minimization problem can be reformulated as the maximization of the surrogate function, leading to the following  problem:
\begin{subequations}
\label{prob:surrogate}
\begin{align}
    \max_{S} & \quad f(S)\\
        \mathrm{s.t.}\quad &  S\in \{S_{\min}, S_{\min}+1,\dots,S_{\max}\},\label{eq:S_range}
\end{align}
\end{subequations}
where $S_{\min} = \max \{ 1, \lceil\frac{B_0-K_{\max}}{B_1}\rceil\}$ denotes the minimum number of selected features constrained by $K_{\max}$, and $S_{\max} = \min \{ \lfloor\frac{B_0-1}{B_1}\rfloor,D\}$ represents the maximum value that guarantees at least one processed observation.

The surrogate $f(S)$ is found to be a concave function of $S$, exhibiting a unique maximum, established in Proposition \ref{prop:optimal_ratio_LR}.

\begin{Proposition}[Optimal Number of Selected Features]
\label{prop:optimal_ratio_LR}
Let 
\begin{align}
\label{eq:nu(x)}
\nu(x)=&P_{\sf{act}} g(x) \left(\hat{G}_2(x)-\frac{1}{2}\hat{G}_1(x)g(x)\right)\nonumber\\
    &+\frac{G_{\sf{th}} ((B_0-B_1x)g^2(x)-2\hat{G}_2(x) )}{2(B_0-B_1x)^2}.
\end{align}
where $g(x)$ is the DG function defined in \eqref{eq:DG_function}, and $\hat{G}_1(x),\hat{G}_2(x)$ are defined in \eqref{eq:sum_DF_function}.
The optimal number of selected features that solves Problem \eqref{prob:surrogate} is then
\begin{equation}
   S^*= 
   \left \lfloor x^* \right\rceil_{ f(\cdot)},
\end{equation}
where the rounding operator $\lfloor x \rceil_{{ f(\cdot)}}$ is equal to $\lfloor x \rfloor$ if $ f(\lfloor x \rfloor)\geq f(\lceil x \rceil) $, and is otherwise equal to $\lceil x \rceil$. 
The value $x^*$ is given by
\begin{equation}
x^*  = \left\{x| \nu(x)=0, x\in [S_{\min},S_{\max}]  \right\},
\end{equation}
if $\nu(S_{\min})\cdot \nu (S_{\max})<0$ holds, otherwise $S^*  = \argmax_{x\in \{S_{\min},S_{\max} \}}f(x)$.
\end{Proposition}
\noindent The proof is provided in Appendix \ref{proof:optimal_ratio_LR}.

Proposition \ref{prop:optimal_ratio_LR} provides an optimal number of selected features for transmission which minimizes the InfOut probability for enhancing reliability.
The optimal selection can be determined by finding the zero of the first derivative of $f(S)$, which is equivalent to solving $\nu(x)=0$. 
The optimal solution can be obtained using a bisection search over the feasible range $ S\in [S_{\min}, S_{\max}]$ with the complexity of $\mathcal{O}(\log \frac{S_{\max}-S_{\min}}{\varepsilon})$ and tolerance $\varepsilon$.

\section{Extension to  CNN Classification}
\label{sec:outage_minimization_CNN}

In this section, we consider the case of CNN classification and analyze the associated InfOut probability by approximating the inference accuracy distribution using the corresponding receive DG. Subsequently, we address the InfOut probability minimization problem by estimating the distribution of the defined receive CNN DG.

\subsection{Approximation for Receive CNN DG}

\begin{figure}[t!]
\centering
\subfigure[PDF of Inference Accuracy]{
\label{fig:pdf_acc}
\includegraphics[width=0.42\columnwidth]{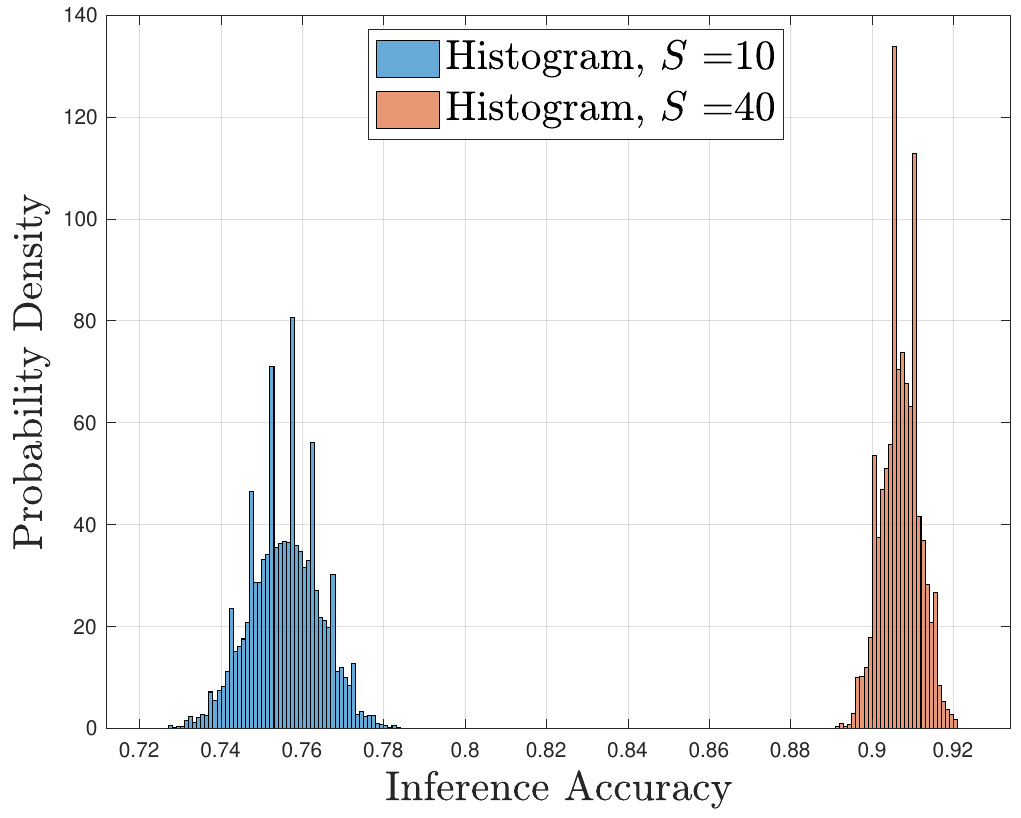}} 
\subfigure[PDF of Receive CNN DG]{
\label{fig:pdf_cnn_dg}
\includegraphics[width=0.42\columnwidth]{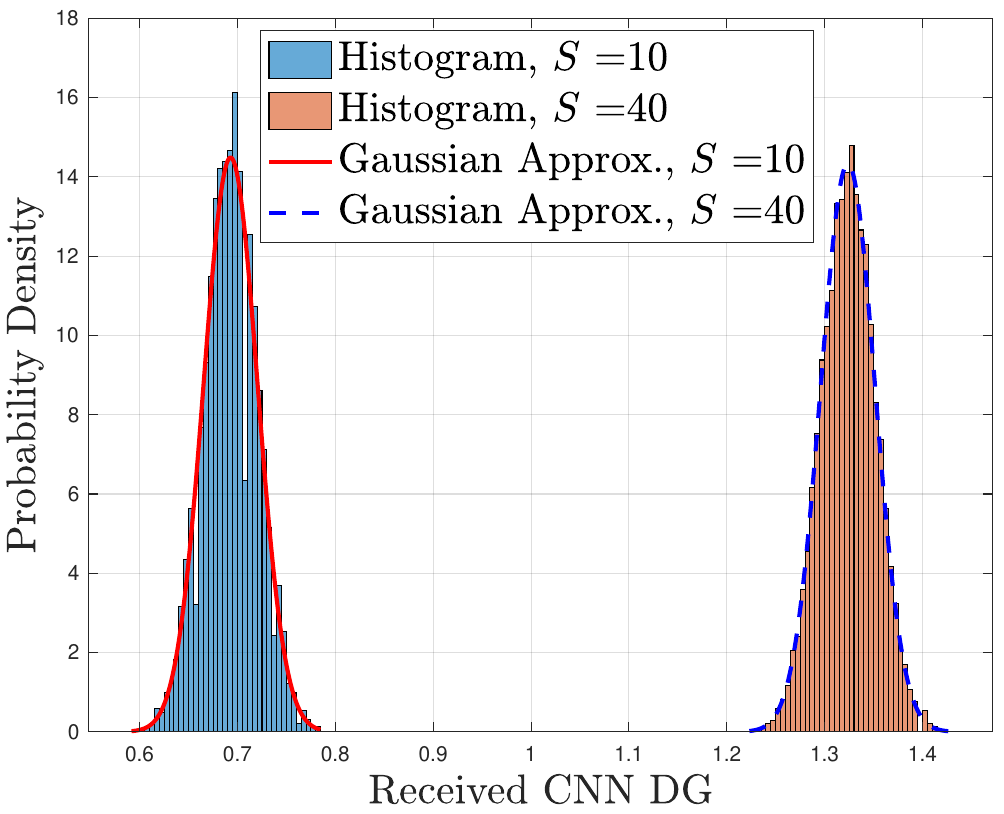}}
\caption{Under the magnitude based feature selection scheme, the distribution comparison between inference accuracy and receive CNN DG using the VGG16 model~\cite{simonyan2015deep} and the ModelNet dataset \cite{ModelNet-Ref}. The settings are $D=512, L=20, \alpha=1, \beta =1, K=12,P_{\sf{act}}=0.7$.\vspace{-3mm}}
\label{fig:CNN_ACC_DG_Approx}
\end{figure}
Unlike linear classifiers, the nonlinearity of the CNN classifier makes it complicated to model the inference accuracy distribution.
As shown in Fig. \ref{fig:pdf_acc}, the PDF of inference accuracy exhibits an intractable distribution that varies with the selected features, making InfOut probability computation challenging.
Moreover, DG based feature selection is not applicable to the CNN case due to the unknown dimension-wise DG of generic CNN feature distributions.
To address these challenges, we leverage the two insights from linear classification, as shown in Remark~\ref{Rem:Insights}.
\begin{Remark}[Insights from Linear Classification]
\label{Rem:Insights}
Two insights from the linear case guide the subsequent CNN extension.
First, under i.i.d.\ feature losses, the receive DG can be written as a sum of many per-feature contributions, and hence admits an accurate Gaussian approximation when the number of retained features is sufficiently large.
Second, under this approximation, the InfOut probability becomes a monotone function of the receive-DG surrogate, which motivates a DG-based feature-selection rule that yields a unique optimum to balance the C$^2$ tradeoff under E2E latency constraint.
\end{Remark}
Building on these findings, we employ a magnitude based feature selection scheme that selects the top-$S$ features with the largest magnitudes, and define the receive CNN DG as follows.
This enables computing the InfOut probability using a tractable distribution.

\begin{Definition}[Receive CNN Discriminant Gain]
\label{Def: CNN DG}
Given the inference accuracy of CNN classifier, denoted as $a_{\sf cnn}$, the corresponding receive CNN DG, $G_{\sf{cnn}}$, is quantified as a monotone increasing function of $a_{\sf cnn}$, given by
\begin{equation}
\label{A2DG}
    G_{\sf{cnn}} = \alpha Q^{-1}(\beta(1-a_{\sf cnn})),
\end{equation}
where $\alpha$ and $\beta$ are the fitting parameters.
\end{Definition}

Using this mapping and the magnitude based feature selection, we approximate the distribution of the receive CNN DG as a Gaussian random variable under a sufficiently large number of selected features (i.e., $S\geq S_{\min}$):
  \begin{equation}
  \label{eq:Gaussian_cnn}
        G_{\sf{cnn}}\sim \mathcal{N} \left(\mu_{(K,S,P_{\sf{act}})}, \sigma^2_{(K,S,P_{\sf{act}})} \right),
    \end{equation}
   where $\mu_{(K,S,P_{\sf{act}})}$ and $\sigma^2_{(K,S,P_{\sf{act}})} $ are the mean and variance of the receive CNN DG, respectively.
   These coefficients are determined by the number of observations $K$, transmitted feature number $S$, and channel activation probability $P_{\sf{act}}$.

In Fig. \ref{fig:CNN_ACC_DG_Approx}, we validate the Gaussian approximation of the receive CNN DG using a real dataset. 
Fig. \ref{fig:pdf_acc} demonstrates an irregular and intractable PDF of inference accuracy.
By using the defined receive CNN DG,
Fig. \ref{fig:pdf_cnn_dg} shows that the Gaussian approximation accurately captures the distribution across different settings of selected features.


\begin{algorithm}[t]
 \caption{Receive CNN DG Distribution Estimation}
 \begin{algorithmic}[1]
 \renewcommand{\algorithmicrequire}{\textbf{Input:}}
  \REQUIRE  Sets of activation probability $\mathcal{P}_{\sf{act}}$, number of observations $ \mathcal{K}$, and selected features $\mathcal{S}$
\STATE{Initialization}: Training datasets and well-trained model;
  \FOR  {Network Parameters: $P_{\sf{act}}\in \mathcal{P}_{\sf{act}},K\in \mathcal{K},S\in \mathcal{S}$ }
  \FOR{Number of trials: $n = \{1,2,\dots,N\}$ }
  \FOR{ Data samples in training dataset}
   \STATE Extract feature vector $\mathbf{x}\in \mathbb{R}^D$ using the selected observation batch with the size of $K$;
   \STATE Compute the channel-effect feature vector $\hat{\mathbf{x}}=[\hat{x}(1),\dots,\hat{x}(D)]$ by 
   randomly masking Top-$S$ features with $P_{\sf{act}}$ and setting others as zeros;
  \STATE Infer label using $\hat{\mathbf{x}}$;
    \ENDFOR
    \STATE Compute the inference accuracy $a_{\sf cnn}(n)$;
    \STATE Compute the receive CNN DG $G_{\sf{cnn}} (n)$ with \eqref{A2DG};
  \ENDFOR
    \STATE Compute the estimated mean of DG $\hat{\mu}_{(K,S,P_{\sf{act}})}= \frac{1}{N}\sum_{n=1}^N G_{\sf{cnn}}(n)$;
    \STATE Compute the estimated  variance  $\hat{\sigma}^2_{(K,S,P_{\sf{act}})}=  \frac{1}{N-1} \sum_{n=1}^N (G_{\sf{cnn}}(n)-\hat{\mu}_{(K,S,P_{\sf{act}})})^2$ ;
  \ENDFOR
  \RETURN Lookup table of receive CNN DG distribution: $\{\hat{\mu}_{(K,S,P_{\sf{act}})}\},\{\hat{\sigma}_{(K,S,P_{\sf{act}})}\}$
 \end{algorithmic} 
 \label{Alg}
 \end{algorithm}

\subsection{Optimal Magnitude based Feature Selection}

Building on the Gaussian approximation of receive CNN DG in~\eqref{eq:Gaussian_cnn}, the InfOut probability of CNN cases under the accuracy threshold $A_{\sf{th}}$ can be expressed as 
    \begin{equation}
\label{eq:outage_prob_cnn}
\begin{split}
   P_{\sf{out}}^{\sf{e2e}} &= \Pr(a_{\sf cnn}\leq A_{\sf{th}}) \\
   & = \Pr\left( G_{\sf{cnn}} \leq G_{\sf{th}} \right)\\
   &= \frac{1}{\sqrt{2 \pi} \sigma_{(K,S,P_{\sf{act}})}}\int_{-\infty}^{G_{\sf{th}}} \exp{\left(-\frac{(x-\mu_{(K,S,P_{\sf{act}})})^2}{2\sigma^2_{(K,S,P_{\sf{act}})}}\right)} d x   \\
   &= \frac{1}{\sqrt{2 \pi}}\int_{-\infty}^{-\Psi_{\sf{cnn}}} \exp{\left(-\frac{y^2}{2}\right)} d y   \\
    &=1- Q\left( -\Psi_{\sf{cnn}} \right) \\ 
     &=Q\left( \Psi_{\sf{cnn}} \right)
    \end{split}
\end{equation}
where $Q(x)=\frac{1}{\sqrt{2\pi}}\int_x^\infty \exp{ \left( -\frac{t^2}{2} \right) dt}$ is the Gaussian Q-function and $\Psi_{\sf{cnn}}$
is the surrogate function that minimizes the InfOut probability of CNN classification,
given by
\begin{equation}
\Psi_{\sf{cnn}}=\frac{\mu_{(K,S,P_{\sf{act}})} - G_{\sf{th}}} {
\sigma_{(K,S,P_{\sf{act}})}   } 
\end{equation}
with $ G_{\sf{th}}= \alpha Q^{-1}(\beta(1-A_{\sf{th}}))$ being the threshold of the required receive CNN DG.
It follows that the InfOut probability in the CNN case is a monotonically decreasing function of the surrogate function $\Psi_{\sf{cnn}}$.

However, maximizing $\Psi_{\sf{cnn}}$
depends on the unknown parameters  $\mu_{(K,S,P_{\sf{act}})}$ and $\sigma_{(K,S,P_{\sf{act}})}$. 
To estimate these parameters, we develop an algorithm that emulates the effects of random feature loss on the inference process using training datasets, as outlined in Algorithm \ref{Alg}.
Using the estimated parameters, the optimization problem for CNN classification is reformulated as 
\begin{equation}
\label{prob:outage_mini_cnn_est}
    \begin{split}
        \max_{S} & \quad  \hat{\Psi}_{\sf{cnn}}(S) \\
        \text{s.t.}  & \quad \eqref{eq:S_range}, \eqref{eq:view-feature relation} ,
    \end{split}
\end{equation}
where $\hat{\Psi}_{\sf{cnn}}(S)$ is the estimated surrogate  expressed in terms of the number of selected features $S$, given by
\begin{equation}
    \hat{\Psi}_{\sf{cnn}}(S)=\frac{\hat{\mu}_{(\hat{K},S,P_{\sf{act}})} - G_{\sf{th}}} { \hat{\sigma}_{(\hat{K},S,P_{\sf{act}})}},
\end{equation}
Here, $\hat{K}$ is the maximum achievable number of processed observations in \eqref{eq:view-feature relation}. $\hat{\mu}_{(\hat{K},S,P_{\sf{act}})} \approx \mu_{(\hat{K},S,P_{\sf{act}})}$  and $\hat{\sigma}_{(\hat{K},S,P_{\sf{act}})}\approx \sigma_{(\hat{K},S,P_{\sf{act}})} $ are the estimated parameters using Algorithm \ref{Alg}.
With knowledge of the long-term CSI (i.e., the distribution of channel gain) at the transmitter, the channel activation probability can be computed using \eqref{active-prob}.
Conditioned on $P_{\sf act}$, the optimal number of selected features is obtained by identifying the solution $S\in \{S_{\min}, \dots, S_{\max}\}$ that maximizes the estimated surrogate function $\hat{\Psi}_{\sf{cnn}}(S)$. This solution can be efficiently found using a bisection search, with the complexity of $\mathcal{O}(\log(S_{\max}- S_{\min}+1))$.

\section{Experimental Results}
\label{sec:experiments}


\subsection{Experimental Setup}
Unless specified otherwise, the default experimental settings  are as follows:

\subsubsection{Computation and communication configuration}

We present an edge inference framework consisting of an edge device and an edge server, operating under a $T_{\max}=10$ ms E2E latency constraint that encompasses both on-device computation and feature transmission.
For the computation settings, the edge device randomly selects $K$ observations of the target object for feature extraction using the VGG16 model, which contains 14.7 million network parameters \cite{simonyan2015deep}. With this feature extractor, the computation workload for extracting features from a single observation is 936.2 MFLOPs. The edge device is equipped with an NVIDIA Jetson TX2 Series, which provides a computation speed of $f_c=1$ TFLOPs/s~\cite{NVIDIA_Jetson_TX2}.
For the communication settings, each feature is quantized to  $Q_B=16$ bits, with the index quantized by $Q_I=9$ bits, and is assumed to be transmitted within $T_\Delta=0.3$ ms. 
Thus, the resulting computation latency is $T_{\sf comp}(K)= 0.936K$ ms, while
 transmission latency is comparable and modeled as $T_{\sf comm}(S)=0.3S$ ms.
The system bandwidth is 
$B_W=5$ MHz, and the noise power spectral density at the receiver is $N_0=10^{-9}$ W/Hz~\cite{QSFL}. The Rayleigh fading channel gain is modeled as $h\sim\mathcal{CN}(0,1)$.
The resulting channel outage probability is given by
\begin{equation}
    P_{\sf{out}} = 1 - \exp{\left( -\frac{N_0B_W}{P_{\max}}\left( 2^{\frac{Q_B+Q_I}{T_\Delta B_W}}-1 \right) \right)},
\end{equation}
which adapts to the transmit power constraint $P_{\max}$.
The accuracy requirements are set at 97\% of the maximum achievable accuracy, resulting in $A_{\sf{th}}=96.8\%$ for linear classification and $A_{\sf{th}}=87.3\%$ for CNN based classification.

\subsubsection{Classifier settings} The two classifiers and their corresponding datasets are detailed as follows.
\begin{itemize}
    \item \textbf{Linear classification on synthetic GMM data:} 
    For the linear classifier, feature vectors are generated according to GMM defined in \eqref{data_dis}.  
    The feature vectors have a  dimensionality of $D=30$. 
    The centroid of one cluster is a vector with all elements equal to $+1$, while the centroid of the other cluster is a vector with all elements equal to $-1$.
    The covariance matrix is given by $\mathbf{C}=\diag\{\frac{2}{3}d+10\},d\in\{1,2,\dots,30\}$, modeling the decreasing DG across dimensions.
Building on the dimension-wise DG computed by \eqref{eq:min_DG}, the top-$S$ features are selected for transmission.
The inferred label is obtained by feeding the received features into the classifier given in \eqref{Maha_min_classifier}.

\item \textbf{CNN based classification on real-world data:}
For the CNN classifier, we utilize the well-known ModelNet dataset \cite{ModelNet-Ref}, which contains multi-view object observations (e.g., a person or a plant), and implement the CNN architecture using the VGG16 model \cite{simonyan2015deep}. The VGG16 model is partitioned into a feature extractor and a classifier network, where the feature extractor runs on the device and the classifier operates on the server, following the approach in \cite{Zhiyan-AirPooling}.
The resulting CNN architecture is trained for average pooling and targets a subset of ModelNet dataset containing $L=20$ popular object classes. To perform feature extraction, the device randomly selects $K$ observations of the same class from the dataset and processes them through the feature extractor. 
Specifically, each ModelNet image is resized from $3\times224 \times224$ to $3\times56 \times56$ before being processed by the on-device feature extractor, producing a $512\times 1 \times 1$ tensor \cite{ZW2024ultra-LoLa}. The top-$S$ elements with the highest amplitudes in the feature tensor are selected for transmission. Finally, the received feature tensor is reconstructed and passed to the server-side classifier to generate the inferred label.
\end{itemize}


\subsubsection{Benchmarking schemes} 
To evaluate the performance of the proposed optimal C$^2$ tradeoff, we consider the following benchmark schemes.
\begin{itemize}
    \item \textbf{Brute-force search:} 
   Given the channel outage probability $ P_{\sf{out}}$, the feasible solution set is determined by identifying all pairs of the number of observations $K$ and selected features $S$ that satisfy the E2E latency constraint. The optimal solution is then obtained through an exhaustive search over all feasible solutions to minimize the InfOut probability.
   
       \item \textbf{Maximal features (MaxFeat):} 
  This communication-dominant approach allocates most of the latency budget to feature transmission. Among the feasible solutions that satisfy the latency constraint, this scheme prioritizes the one with the maximum number of features. Once the solution with the maximum features is identified, the number of observations is maximized.
    \item \textbf{Maximal observations (MaxObs):} 
  Unlike MaxFeat, this scheme focuses on incorporating as many observations as possible by extending the computation latency. After identifying feasible solutions with the maximum number of observations, the number of features is maximized.

\item \textbf{Accuracy-threshold based maximal observations/features (ATB-MaxObs/ATB-MaxFeat):}
Unlike the previous baselines, which only maximize the number of observations or selected features, this scheme enforces an accuracy requirement on feasible solutions---a common practice in conventional edge inference under the assumption of reliable channels~\cite{li_zeng_zhou_chen_2020,Zhiyan-JASC-2023,RN291}. Specifically, it uses one-shot inference on the training dataset to filter out solutions whose predicted accuracy falls below a prescribed threshold, and then selects the final solution by maximizing either the number of observations (ATB-MaxObs) or the number of selected features (ATB-MaxFeat), as described above.
Mathematically, let $a_{\sf pred}(K,S)$ denote the profiled one-shot accuracy under reliable channels ($P_{\sf act}=1$). The predicted-feasible set is
$\mathcal{F}_{\sf pred}
=
\Big\{(K,S)\ \big|\ 
a_{\sf pred}(K,S)\ge A_{\sf th},\ 
\frac{K N_F}{f_c}+T_\Delta S \le T_{\max}
\Big\}.$
Under fading ($P_{\sf act}<1$), the realized accuracy becomes random; denote it by $a(K,S;P_{\sf act})$.
We quantify the miscalibration of the one-shot accuracy prediction by the \emph{mismatch probability}, given by
\begin{equation}
\varepsilon_b
\triangleq
\Pr\!\left(a(K_b,S_b;P_{\sf act}) \le A_{\sf th}\right),
 (K_b,S_b)\in\mathcal{F}_{\sf pred},
\end{equation}
which measures how often the baseline decision $(K_b,S_b)$ violates the target accuracy under fading. Since the event ${a(K_b,S_b;P_{\sf act})\le A_{\sf th}}$ coincides with the InfOut event for the fixed design $(K_b,S_b)$, the mismatch probability can be computed via \eqref{eq:outage_prob} for linear classification and via \eqref{eq:outage_prob_cnn} for CNN classification.
This implies that $\varepsilon_b$ is monotone increasing with the channel-outage probability (in the practical regime $P_{\sf out}\le 0.5$).

\end{itemize}


\subsection{Computation-communication Tradeoff}

\begin{figure}[t]
\centering
\subfigure[Linear Classification]{
\includegraphics[width=0.46\columnwidth]{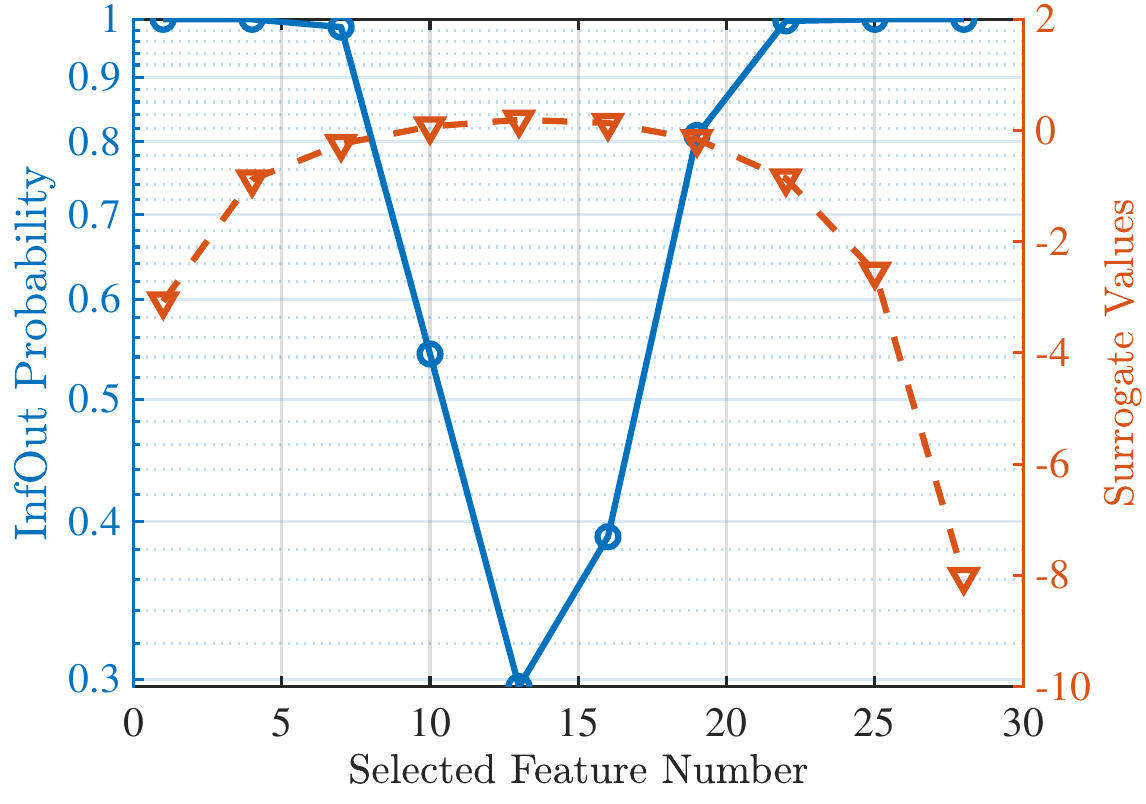}}
\subfigure[CNN Classification]{
\includegraphics[width=0.46\columnwidth]{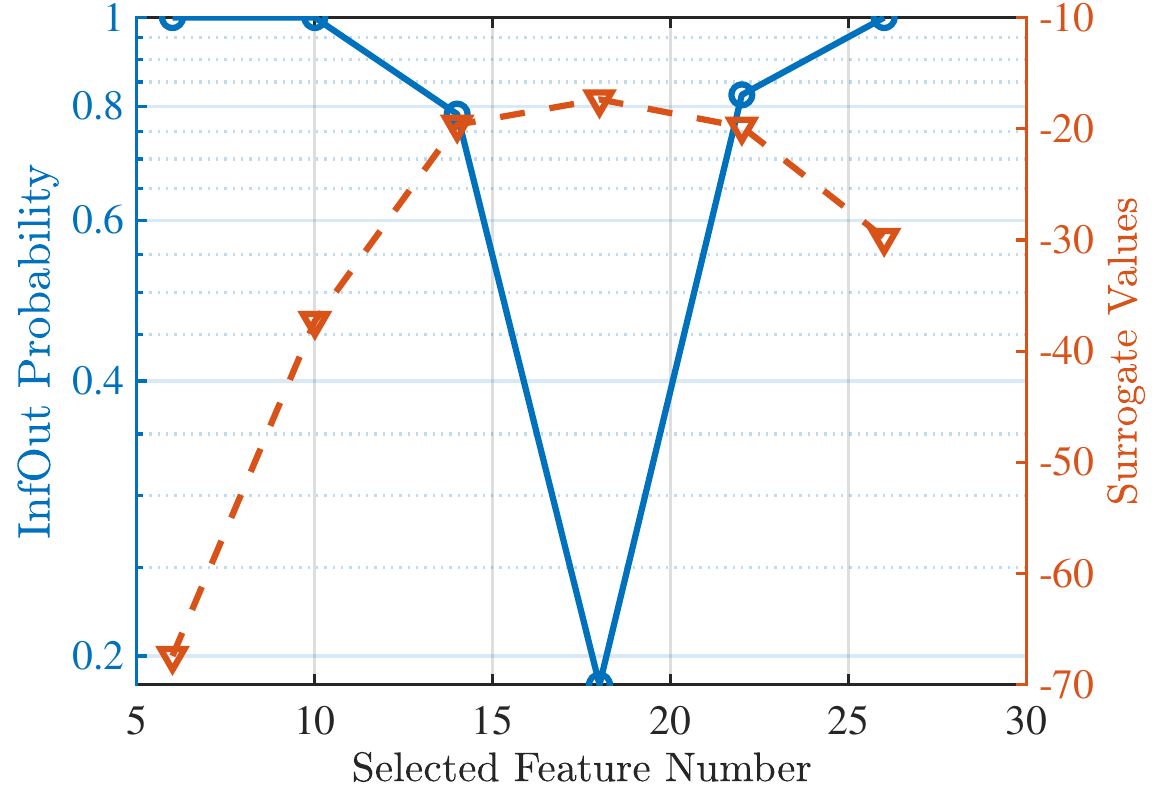}}
\caption{The illustration of the C$^2$ tradeoff under E2E latency constraint, where the relationship between the number of features and observations is modeled by \eqref{eq:view-feature relation}.  \vspace{-2mm}}
\label{fig:tradeoff_demo}
\end{figure}

The C$^2$ tradeoff is illustrated in Fig. \ref{fig:tradeoff_demo} using the criteria of InfOut probability and surrogate values.
First, as the number of selected features increases, the InfOut probability initially decreases before increasing again, resulting in a unique minimum. 
This behavior illustrates the tradeoff between communication and computation in latency-constrained edge inference systems. Transmitting more features, which increases communication latency, improves the system's ability to withstand channel outages, thereby reducing the InfOut probability. However, the resulting decrease in the number of processed observations eventually degrades feature quality, causing the InfOut probability to rise.
The unimodal nature of the InfOut probability with respect to the number of transmitted features confirms the existence of a unique minimum, as established in Proposition \ref{prop:optimal_ratio_LR}.
Second, the surrogate values exhibit a trend opposite to that of the InfOut probability. This observation validates the findings in \eqref{eq:outage_prob1} and \eqref{eq:outage_prob_cnn} for both linear and CNN based classification models. 
Specifically, both of them reach their extrema at the same point corresponding to the number of selected features.

\begin{figure}[t]
\centering
\subfigure[Linear Classification]{\label{fig:top-1_tradeoff-linear}
\includegraphics[width=0.46\columnwidth]{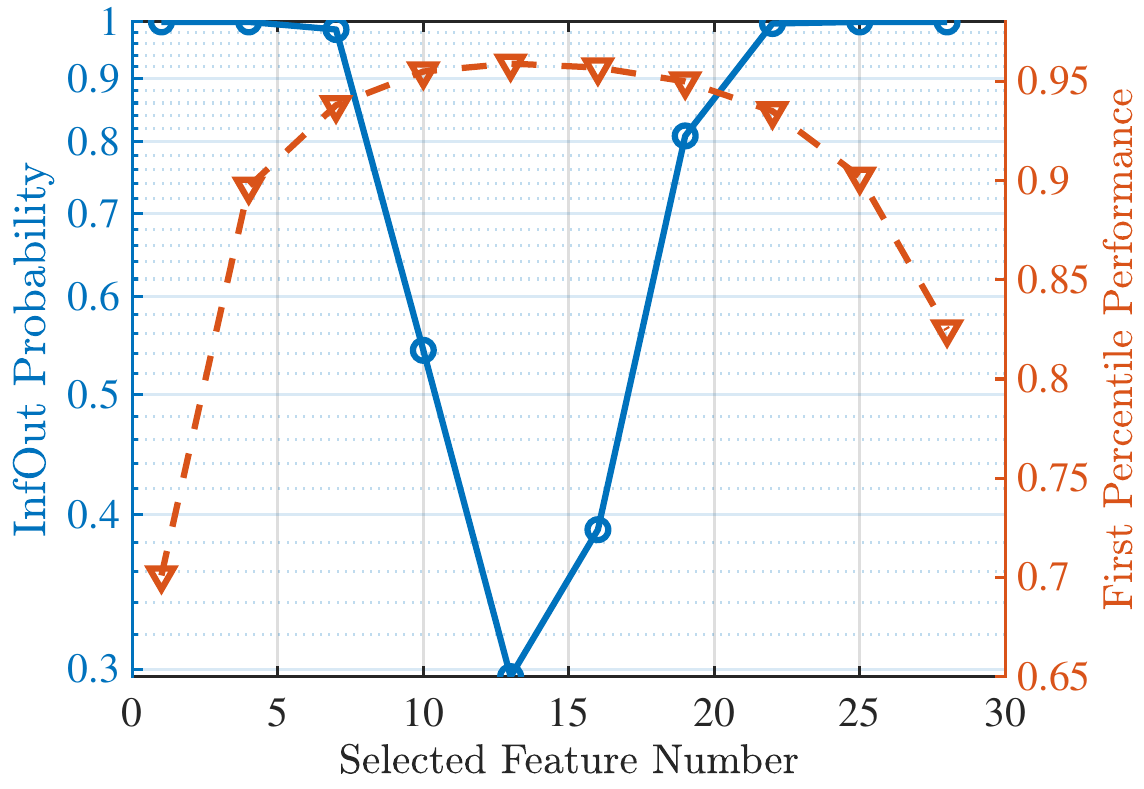}}
\subfigure[CNN Classification]{
\label{fig:top-1_tradeoff-cnn}
\includegraphics[width=0.46\columnwidth]{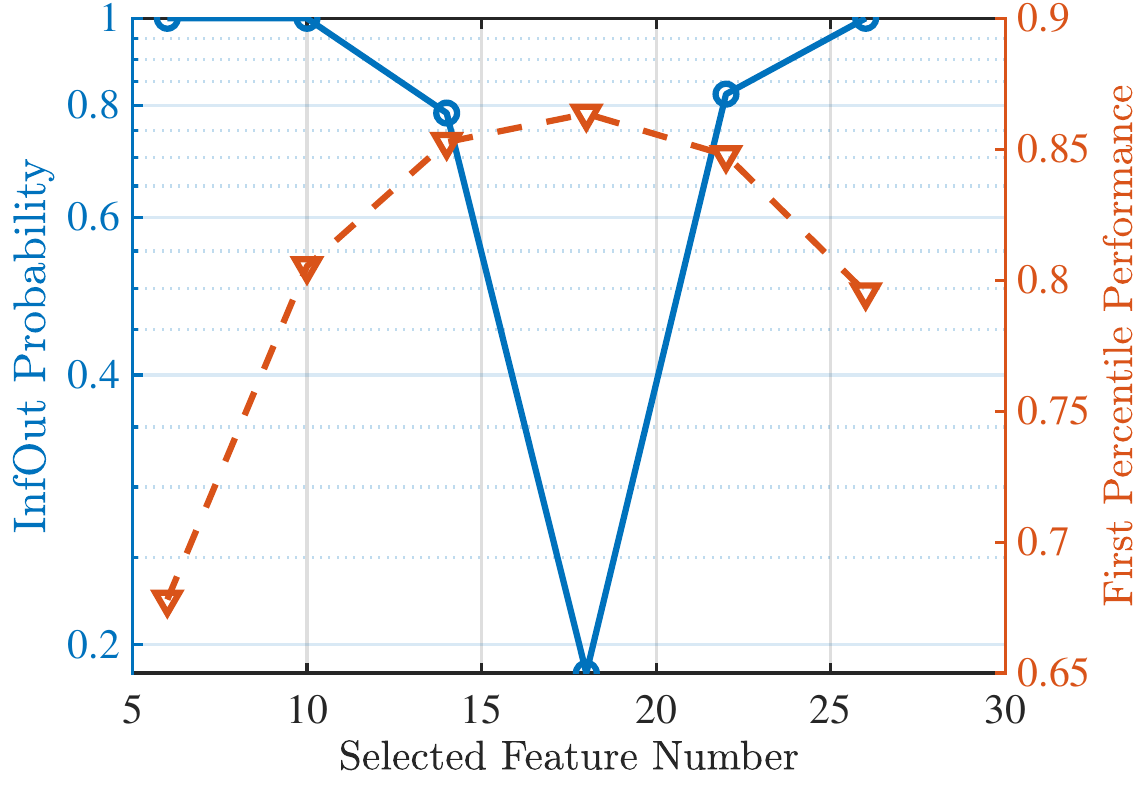}}
\caption{The comparison between InfOut probability and first percentile performance adopted by \cite{RN366}.\vspace{-2mm}}
\label{fig:tradeoff_demo2}
\end{figure}

In Fig. \ref{fig:tradeoff_demo2}, we compare the defined InfOut probability with first percentile performance, which is defined as the value that separates the lowest $1\%$ of samples from the highest $99\%$ of the samples in the accuracy distribution~\cite{RN366}.
As the number of transmitted features increases, first percentile performance exhibits an inverse trend to the InfOut probability, indicating that the defined InfOut probability effectively captures the reliability of edge inference systems.
These observations further validate the tractability and accuracy of the proposed analytical framework, which is derived based on the surrogate function.

\subsection{Inference Outage Performance}

\begin{figure}[t]
\centering
\subfigure[Linear Classification]{
\includegraphics[width=0.42\columnwidth]{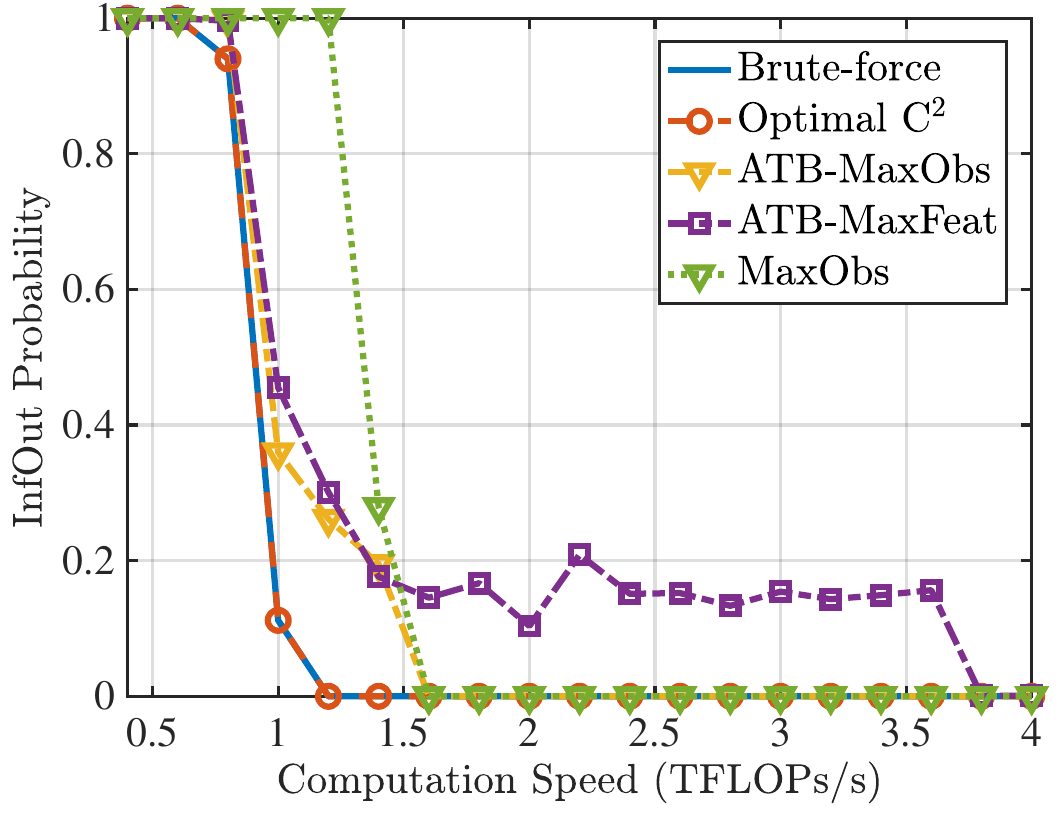}}
\subfigure[CNN Classification]{
\includegraphics[width=0.42\columnwidth]{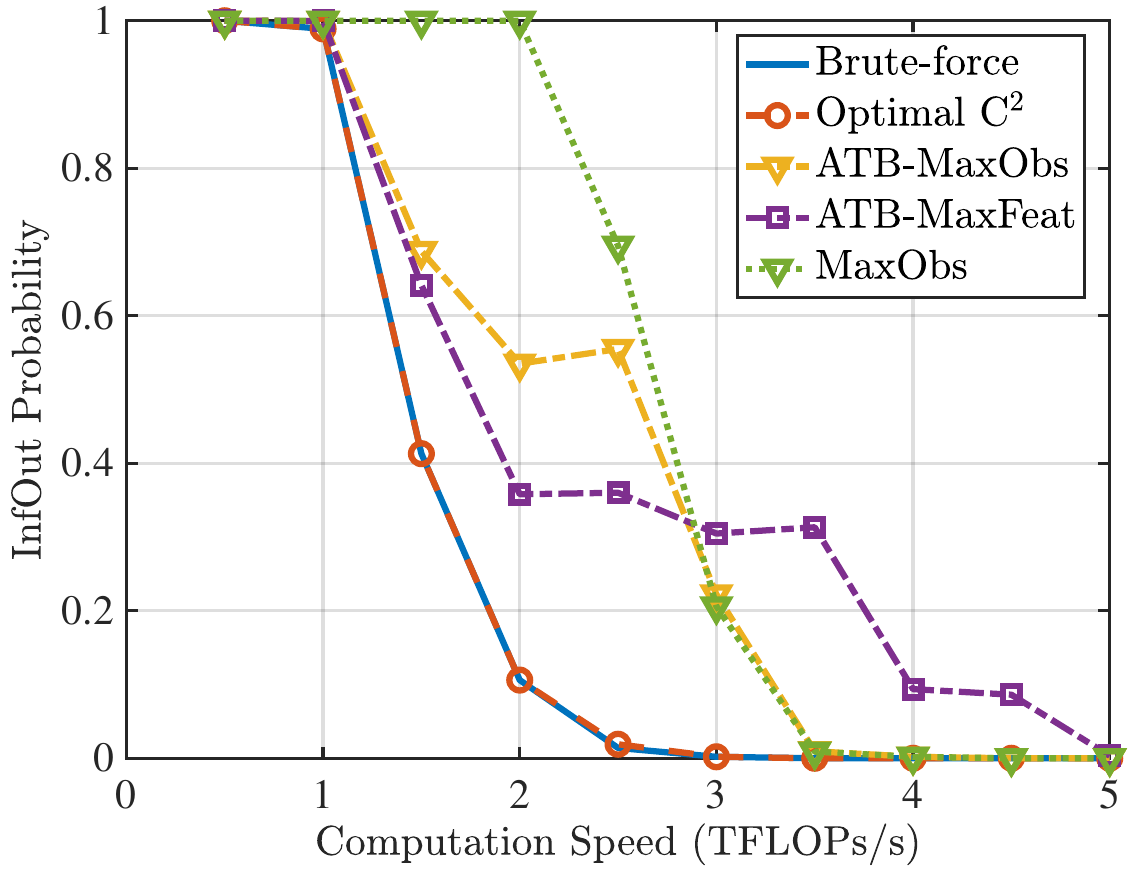}}
\caption{Comparison between optimal C$^2$ scheme and benchmarks for different settings of computation speed.\vspace{-2mm}}
\label{fig:SpeedVsOutage}
\end{figure}

In this subsection, we evaluate inference outage performance by comparing the proposed approach with benchmarks for both linear and CNN based classification.
As shown in Fig. \ref{fig:SpeedVsOutage}, the InfOut probability is evaluated with different computation speeds.
Specifically, the InfOut probability decreases with increasing computation speed across all schemes. 
This behavior is attributed to faster computation, which either enables the processing of more observations during feature extraction (improving feature quality) or allows additional latency to be allocated for transmitting more features (enhancing feature quantity). 
Both factors work together to reduce the InfOut probability.
Furthermore, the proposed scheme consistently outperforms benchmarks such as ATB-MaxFeat, ATB-MaxObs, and MaxObs. 
This superior performance arises from the precise optimization of the C$^2$ tradeoff.
In contrast, ATB-MaxFeat and ATB-MaxObs benchmarks that assume reliable channels and enforce a target accuracy exhibit degraded performance, as they fail to account for the inference accuracy distribution affected by fading channels. The MaxObs scheme prioritizes the number of observations, resulting in significant performance improvements with increased computation speed.
At low computation speed, MaxObs spends most of the latency budget on computation and leaves too few features to transmit. Consequently, the normalized receive DG enters the steep region of the Q-function, causing the InfOut probability to increase sharply.
However, it suffers from severe performance degradation under low computation capacity.
Moreover, the proposed scheme closely approximates the brute-force solution, demonstrating its near-optimal performance. 

\begin{figure}[t]
\centering
\subfigure[Linear Classification]{
\includegraphics[width=0.42\columnwidth]{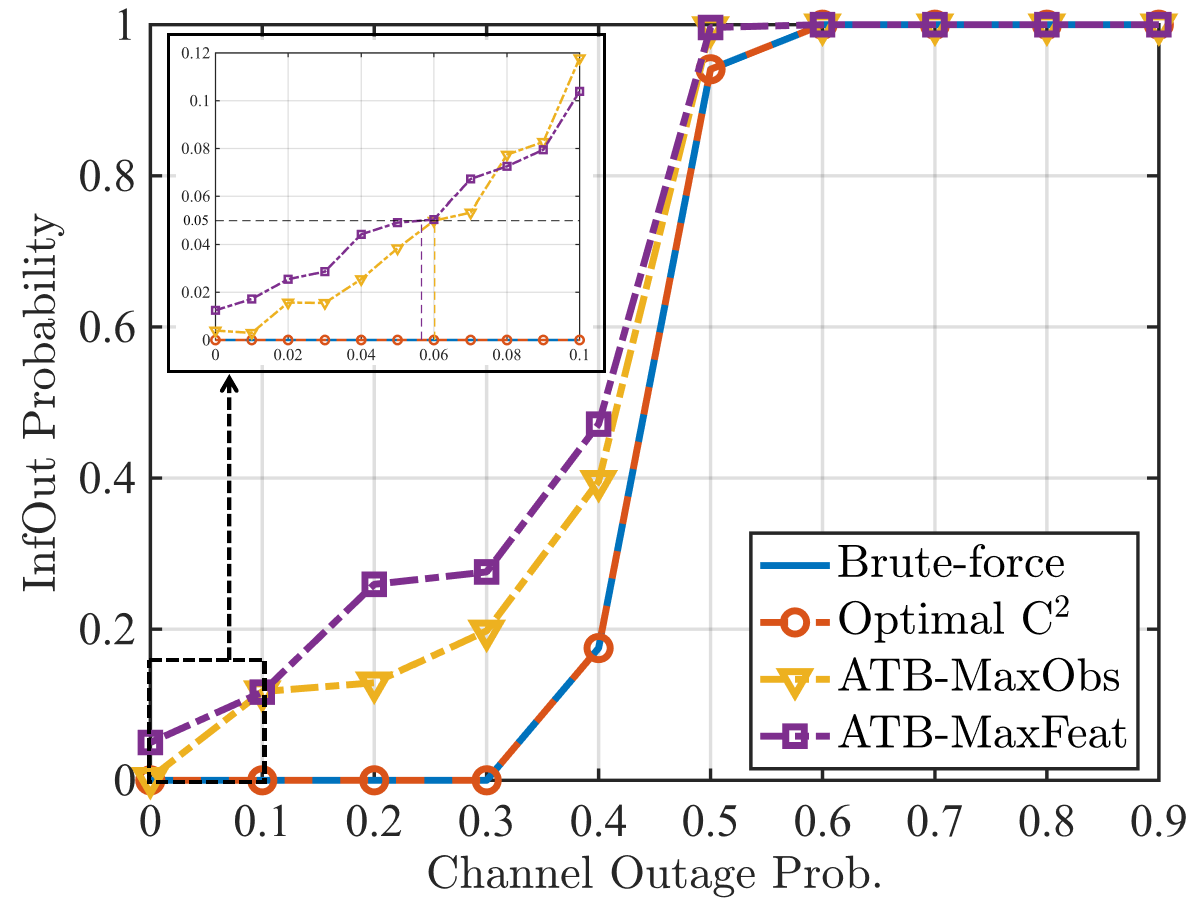}}
\subfigure[CNN Classification]{
\includegraphics[width=0.42\columnwidth]{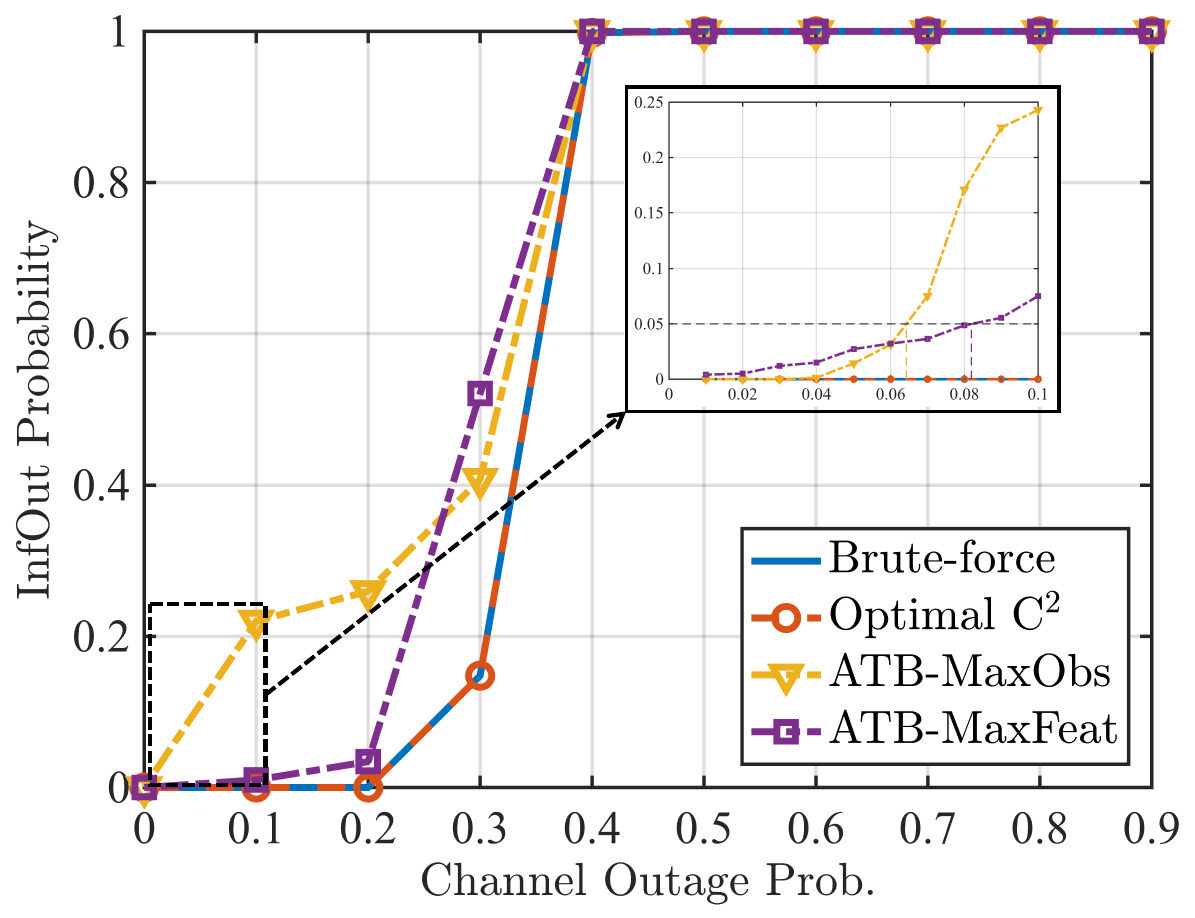}}
\caption{Comparison between optimal C$^2$ scheme and benchmarks for different settings of channel outage probability.\vspace{-2mm} }
\label{fig:ActVsOutage}
\end{figure}

\begin{figure}[t]
\centering
\subfigure[Linear Classification]{
\includegraphics[width=0.42\columnwidth]{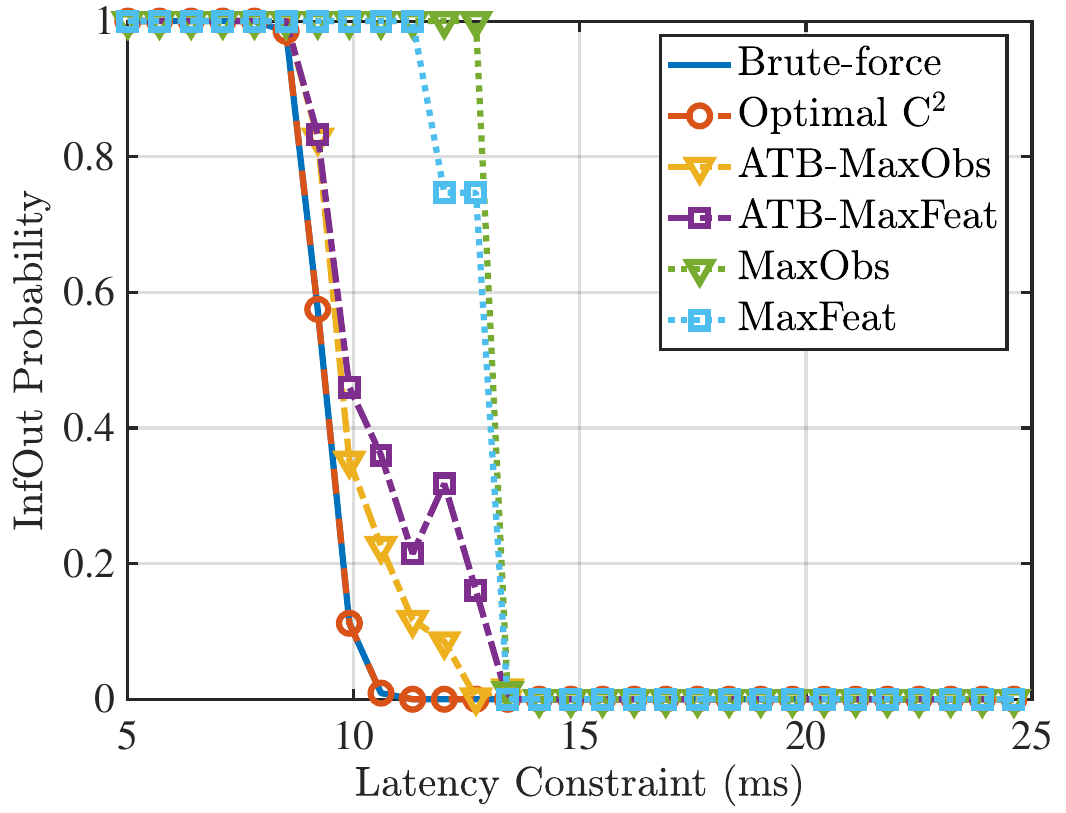}}
\subfigure[CNN Classification]{
\includegraphics[width=0.42\columnwidth]{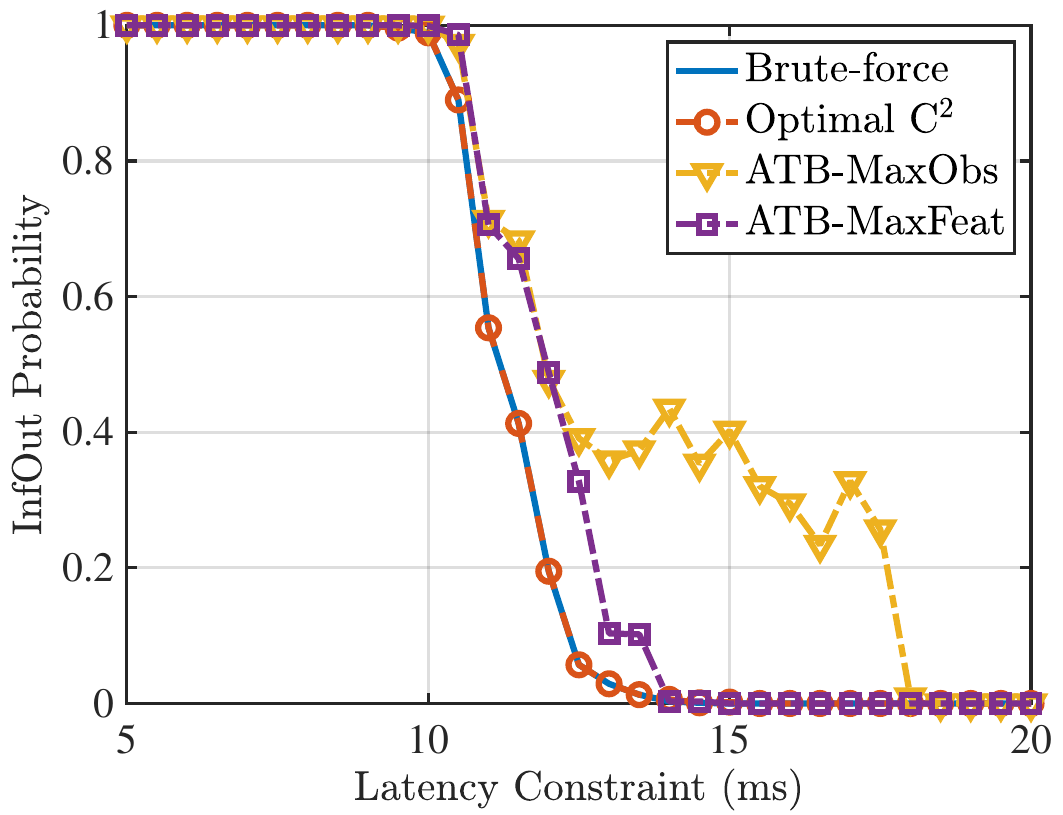}}
\caption{Comparison between optimal C$^2$ scheme and benchmarks for different settings of latency constraints.\vspace{-2mm}}
\label{fig:LatencyVsOutage}
\end{figure}

\begin{figure}[t]
\centering
\subfigure[Linear Classification]{
\includegraphics[width=0.42\columnwidth]{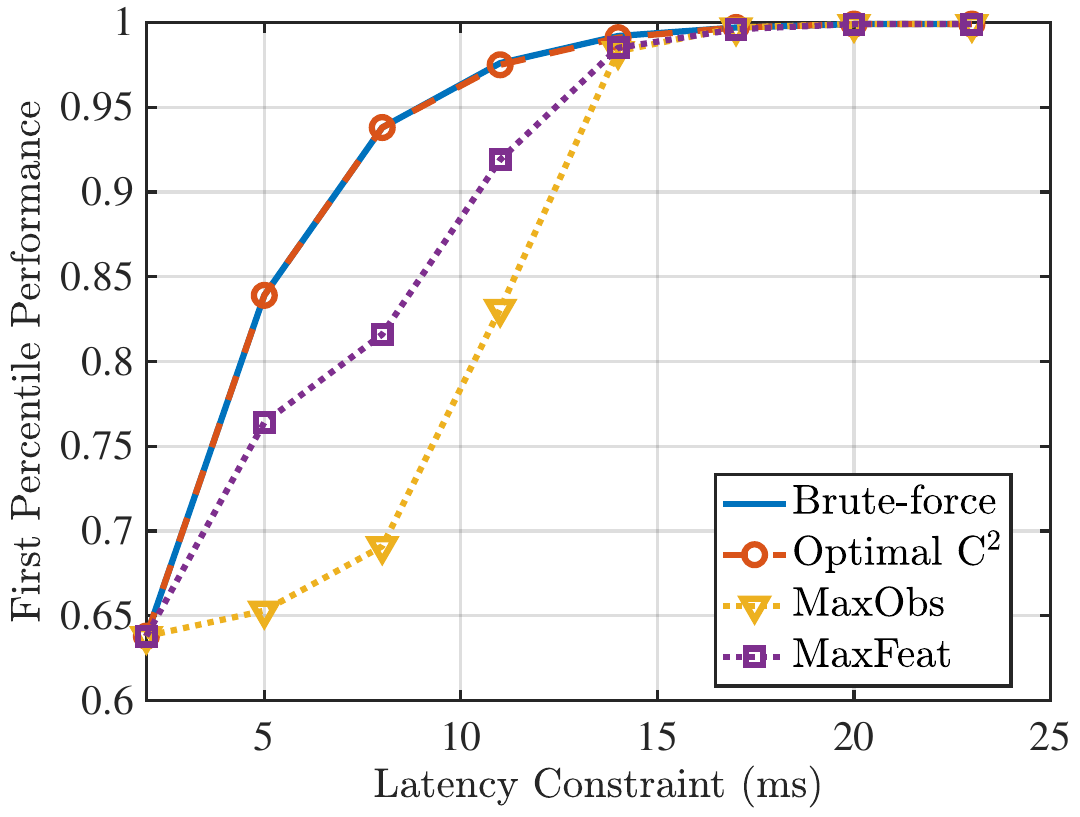}}
\subfigure[CNN Classification]{
\includegraphics[width=0.42\columnwidth]{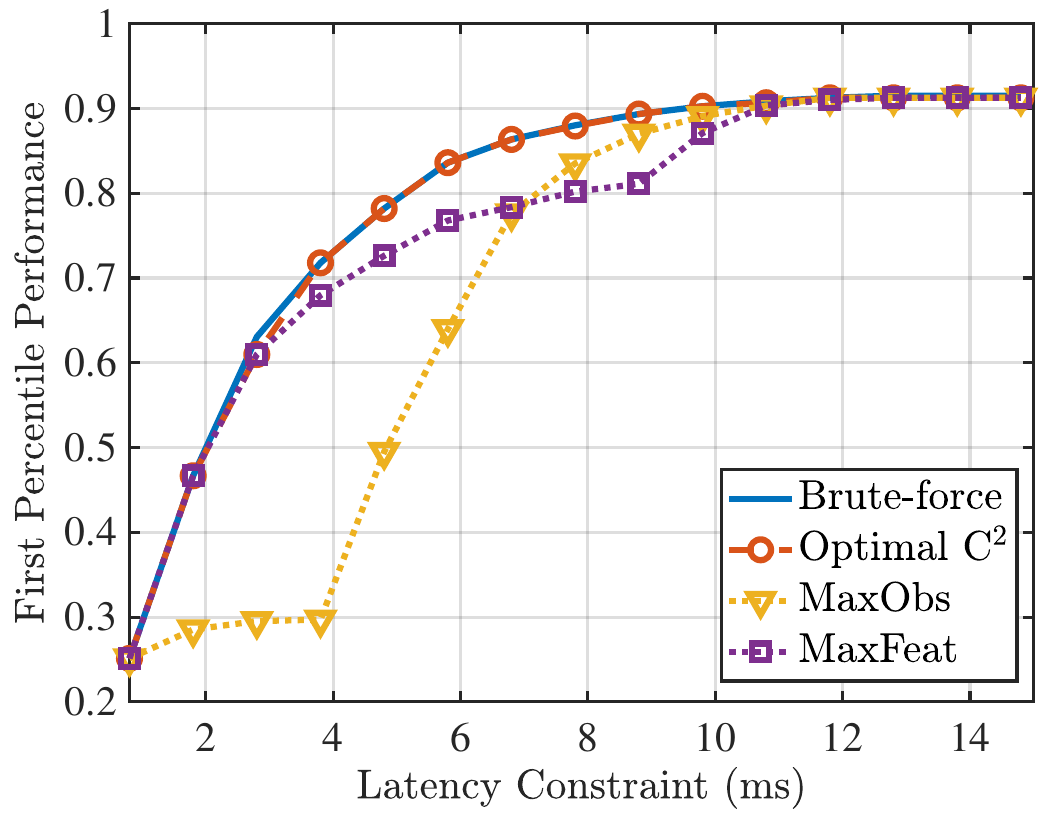}}
\caption{The first percentile performance comparison between optimal C$^2$ scheme and benchmarks for different settings of latency constraints. \vspace{-3mm}}
\label{fig:LatencyVsKPP}
\end{figure}

Then, we evaluate the effects of channel outage probability on system performance, as shown in Fig. \ref{fig:ActVsOutage}.
As the channel outage probability increases, the InfOut probability correspondingly rises. 
A higher channel outage probability results in fewer features being received by the edge server, thereby compromising inference system reliability.
Within the channel outage probability range of 
$[0,0.5]$ for the linear classifier and 
$[0,0.4]$ for CNN based classification, the proposed scheme maintains a remarkably low InfOut probability, consistently outperforming the benchmark methods. This underscores the advantage of optimizing the C$^2$ tradeoff in enhancing the reliability of latency-constrained edge inference systems.
However, when the channel outage probability exceeds 0.5, none of the schemes can guarantee reliable inference.
The reason is that, when nearly half of the transmitted features are erased, meeting the target accuracy becomes unlikely, pushing the InfOut probability close to one.
In addition, to operate in a rare-outage (high-reliability) regime with \(P_{\sf out}^{\sf e2e}\leq 0.05\), the channel-outage probability of the baselines must be limited to \(P_{\sf out}\leq 0.06\) for linear classification and \(P_{\sf out}\leq 0.08\) for CNN classification. By contrast, the proposed approach maintains \(P_{\sf out}^{\sf e2e}\leq 0.05\) up to \(P_{\sf out}\leq 0.3\) for the linear case and \(P_{\sf out}\leq 0.2\) for CNN classification. This improved tolerance to channel outages stems from the optimized C$^2$ tradeoff with respect to the channel state.

Next, we compare the proposed scheme with benchmarks under various latency constraints in Fig. \ref{fig:LatencyVsOutage}.
The results show that a more relaxed latency constraint leads to a lower InfOut probability. 
This is because a larger E2E latency allows for processing more observations and transmitting additional features, thereby improving both feature quality and quantity, which in turn reduces the InfOut probability.
Similar to the findings in Fig. \ref{fig:SpeedVsOutage}, the proposed scheme consistently demonstrates its optimality across different latency constraints, outperforming the benchmark methods.

Finally, Fig. \ref{fig:LatencyVsKPP} evaluates the worst-case performance of the optimal C$^2$ approach using first percentile performance, as applied in \cite{RN366}.
The results show that first percentile performance remains low across all schemes under stringent latency constraints. As the latency constraint is relaxed, performance improves and eventually saturates as more features are transmitted and more observations are processed, thereby enhancing the reliability of the inference.
Furthermore, the proposed C$^2$ scheme outperforms the benchmarks and closely approaches the brute-force solution across various latency constraint settings.


\section{Conclusion}
\label{sec:conclusion}

We have investigated the reliability of latency-constrained edge inference systems by introducing the concept called InfOut probability. A fundamental C$^2$ tradeoff was revealed and quantified: mitigating inference outage requires the edge device to process more observations and upload more features, which increases both computation and communication overhead. However, these requirements conflict under a constraint on latency. To optimize this tradeoff, we have derived a unimodal surrogate function for linear classification and utilized the insights to achieve near-optimal performance in the case of CNN based classification.

This work represents arguably the first theoretical framework for analyzing the reliability of an edge inference system from the perspective of outage probability.
The revisiting of outage in the new context
opens new directions in communication theory, particularly for edge-AI applications that are both latency-sensitive and mission-critical.
Future research could further investigate the optimal balance among latency, reliability, and inference accuracy by exploring diverse communication techniques such as short-packet transmission, MIMO beamforming, and multi-user resource allocation.



\appendix

\subsection{Proof of Lemma \ref{Lemma:Lindeberg-con}}
\label{proof:Lindeberg-con}

To confirm the satisfaction of Lindeberg’s condition, we verify the \emph{uniform asymptotic negligibility} (UAN) condition as follows:
\begin{equation}
    \lim_{S\rightarrow \infty} \max_{1\leq d \leq S} \frac{\text{Var}(X_d)}{\sigma^2_{\sf{G}}(S)}=\lim_{S\rightarrow \infty} \frac{P_{\sf act}(1-P_{\sf act})\hat{W}_1^2 }{\sigma^2_{\sf{G}}(S)}=0.
\end{equation}

Next, we analyze the Lindeberg term. The upper bound of $\forall \left|X_d-\mathbb{E}[X_d]\right|$ is given by:
\begin{equation}
\begin{split}
     \left|X_d-\mathbb{E}[X_d]\right| & \leq  \max\{ 1-P_{\sf act}, P_{\sf act}\} \hat{W}_d\\
     &\leq  C \triangleq   \max\{ 1-P_{\sf act}, P_{\sf act}\} \hat{W}_1,
\end{split}
\end{equation}
where $C$ denotes the maximum deviation of $X_d$ from its mean. As $S \rightarrow \infty$, we have $\epsilon \sigma^2_{\sf G}(S)=\epsilon P_{\sf act}(1-P_{\sf act}) \sum_{d=1}^S \hat{W}^2_d >C$. Consequently, the indicator function $\mathbb{I}{\left(\left|X_d-\mathbb{E}[X_d]\right|>\epsilon \sigma_{\sf G}(S)\right)}$ equals zero, ensuring that the Lindeberg term vanishes. Thus, Lindeberg’s condition in \eqref{eq:Lindeberg-con} is satisfied, completing the proof.

\subsection{Proof of the Monotonicity of $G_{\sf f}(S)$}
\label{Proof:Monotone_G_F}

Since the dimension-wise DG follows a decreasing order, i.e., $\hat{W}_{d}\geq \hat{W}_{d+1}$, the monotonicity of $ G_{\sf{f}}(S) $ is determined by the sign of the difference $G_{\sf{f}}^2(S+1)-G_{\sf{f}}^2(S)$, given by:
\begin{equation}
\begin{split}
    G_{\sf{f}}^2(S+1)-G_{\sf{f}}^2(S) 
    = & \frac{(G_1(S)+\hat{W}_{S+1})^2}{G_2(S)+\hat{W}_{S+1}^2}-\frac{G_1^2(S)}{G_2(S)}\\
    = &\frac{ \hat{W}_{S+1} G_\Delta(S)}{(G_2(S)+\hat{W}_{S+1}^2)G_2(S)  },
\end{split}
\end{equation}
where $G_\Delta(S) =G_2(S)\hat{W}_{S+1}+2G_1(S)G_2(S) -G_1^2(S)\hat{W}_{S+1}$ is proven to be positive, given by 
\begin{equation}
\begin{split}
      G_\Delta(S)  = & G_2(S)\hat{W}_{S+1}+2G_1(S)G_2(S)-G_1^2(S)\hat{W}_{S+1}\\
                   = & \hat{W}_{S+1}(G_2(S)-G_1^2(S))+2G_1(S)G_2(S)\\
                   = & -\hat{W}_{S+1} \sum_{d_1\neq d_2}\hat{W}_{d_1}\hat{W}_{d_2} + 2\sum_{d_1=1}^S\sum_{d_2=1}^S \hat{W}_{d_1}\hat{W}_{d_2}^2\\
                   \geq & -\hat{W}_{S+1} \sum_{d_1\neq d_2}\hat{W}_{d_1}\hat{W}_{d_2} + \sum_{d_1\neq d_2}  \hat{W}_{d_1}\hat{W}_{d_2}^2+ \sum_{d=1}^S \hat{W}_{d}^3\\
                  \geq &\underbrace{-\hat{W}_{S+1} \sum_{d_1\neq d_2}\hat{W}_{d_1}\hat{W}_{d_2} + \hat{W}_{S+1}\sum_{d_1\neq d_2}  \hat{W}_{d_1}\hat{W}_{d_2}}_{=0} \\
                 &+ \sum_{d=1}^S \hat{W}_{d}^3\\
                 \geq & 0.\\
\end{split}
\end{equation}

Since $G_\Delta(S) \geq 0$, it follows that $G_{\sf{f}}^2(S+1)-G_{\sf{f}}^2(S) \geq 0$ for all $S \in {1,2, \dots, D}$, proving that $G_{\sf{f}}(S)$ is a monotonic increasing function. This completes the proof.

\begin{figure*}[t]
\centering

\begin{equation}
\label{eq;y''<0}
\begin{split}
      y''(x)  & =  \hat{G}_2^{-\frac{3}{2}}(x) \left(\underbrace{(-B_1g^2(x)+(B_0-B_1x)g(x)g'(x) )\hat{G}_2(x)}_{\leq 0} 
        \underbrace{-\frac{(B_0-B_1x)g^4(x)}{4}}_{\leq 0}
    \right)\leq 0.
\end{split}
\end{equation}
\hrulefill
\vspace{-2mm}
\end{figure*}

\subsection{Proof of Proposition \ref{prop:optimal_ratio_LR}}
\label{proof:optimal_ratio_LR}

To simplify notation, we express the surrogate function as a linear combination of two continuous functions over $x\in [0,D]$, given by
\begin{equation}
\label{eq:f(x)}
f(x)=  P_{\sf{act}}f_{1}(x) + f_{2}(x),
\end{equation}
where:
\begin{equation}
\begin{split}
     f_{1}(x)=\frac{\hat{G}_1(x)}{\sqrt{\hat{G}_2(x)}}, \quad
     f_2(x)  =-\frac{G_{\sf{th}}}{(B_0-B_1 x)\sqrt{\hat{G}_2(x)}}.
\end{split}
\end{equation}

Here, the DG based feature selection scheme ensures  several properties of $\hat{G}_1(x)$ and $\hat{G}_2(x)$, given by
   \begin{equation}
       \begin{split}
            \label{eq:dg-Gx}
    \hat{G}_1'(x)=  g(x)  \geq 0,& \quad  \hat{G}'_2(x)  = g^2(x) \geq 0,  \\
    \hat{G}''_1(x) =  g'(x)  \leq 0,& \quad \hat{G}''_2(x)  = 2g(x)g'(x) \leq 0. 
       \end{split}
   \end{equation}

Based on  \eqref{eq:dg-Gx}, we show the concavity of $f(x)$ by separately showing $f_{1}(x)$ and $f_{2}(x)$ are concave functions.
First, we prove that the $f_{1}(x)$ is a concave function.
The first derivative of $f_{1}(x)$ is given by
\begin{equation}
\begin{split}
     f'_{1}(x)
     &=\frac{\hat{G}'_1(x)}{\sqrt{\hat{G}_2(x)}}-\frac{\hat{G}_1(x)\hat{G}'_2(x)}{2\hat{G}^{\frac{3}{2}}_2(x)}\\
    &= \frac{\hat{G}_2(x)g(x)-\frac{1}{2}\hat{G}_1(x)g^2(x)}{\hat{G}^{\frac{3}{2}}_2(x)}.
\end{split} 
\end{equation}

The second derivative of $f_1(x)$ is upper bounded by
\begin{equation}
\begin{split}
       f''_1(x) = 
       & \hat{G}_2^{-\frac{5}{2}}(x)\left( g'(x)\hat{G}_2^2(x)-\hat{G}_1(x)\hat{G}_2(x)g(x)g'(x) \right.\\
       & +\left.   \frac{3}{4}\hat{G}_1(x)g^4(x)-\hat{G}_2(x)g^3(x)\right)\\
        = & \hat{G}_2^{-\frac{5}{2}}(x)\left\{ g'(x)\hat{G}_2(x)[\hat{G}_2(x)-\hat{G}_1(x)g(x)]-g^3(x) \right. \\
        &\times \left. \left[\hat{G}_2(x)-\frac{3}{4}\hat{G}_1(x)g(x)  \right]\right\}\\
        \leq & \hat{G}_2^{-\frac{5}{2}}(x)\left\{ g'(x)\hat{G}_2(x)[\hat{G}_2(x)-\hat{G}_1(x)g(x)]\right.\\
        &\left.-g^3(x)\left[\hat{G}_2(x)-\hat{G}_1(x)g(x)  \right]\right\} \\
        = &\underbrace{\hat{G}_2^{-\frac{5}{2}}(x)}_{\geq 0} \underbrace{[ g'(x)\hat{G}_2(x)-g^3(x)]}_{\leq 0} \underbrace{[\hat{G}_2(x)-\hat{G}_1(x)g(x)]}_{\triangleq \zeta(x) \geq 0}\\
       \leq & 0,\\
\end{split}
\end{equation}
where $\zeta(x)=\hat{G}_2(x)-\hat{G}_1(x)g(x)$ can be proven to be a non-negative function over $x\in [0,D]$. This follows from the fact that its derivative satisfies
\begin{equation}
    \zeta'(x) = -g'(x)\hat{G}_1(x) \geq 0.
\end{equation}
Since $\zeta(x)$ is non-decreasing, its minimum occurs at $x=0$, where
\begin{equation}
  \zeta(x) \geq   \min_x \zeta(x) =\zeta (0)= \hat{G}_2(0)-\hat{G}_1(0)g(0)=0,
\end{equation}
where $\hat{G}_2(0)=\hat{G}_1(0)=0$.
Thus, $\zeta(x) \geq 0$ for all $x$, confirming that $f_1(x)$ is concave as $f''_1(x) \leq 0$.

Next, we show that $f_2(x) = -\frac{G_{\sf{th}}}{y(x)}$ is concave. Let 
\begin{equation}
    y(x)=(B_0-B_1 x)\sqrt{\hat{G}_2(x)}.
\end{equation}

The second derivative of $f_2(x)$ is given by:
\begin{equation}
\label{eq:f''_2(x)}
    f''_2(x) = \frac{G_{\sf{th}}}{y^3(x)} (y''(x)y^2(x)-2(y'(x))^2).
\end{equation}

From \eqref{eq:f''_2(x)}, it suggests that $y''(x)\leq 0$  is a sufficient condition that $f_2(x)$ is a concave function (i.e., $ f''_2(x)\leq 0 $), which can be proven by \eqref{eq;y''<0}.
In the end, $f(x)$ is a concave function of $x$ due to the sum of two concave functions.

Based on the analysis above, the resulting optimal solution ensuring the maximum of $f(x)$ can be obtained by finding the zero of $f'(x)=0$, given as
\begin{equation}
     x^*=\{x\mid f'(x)=0, x\in [S_{\min}, S_{\max}] \},
\end{equation}
where the derivative $ f'(x)$ is given by:
\begin{equation}
    f'(x) = \hat{G}_2^{-\frac{3}{2}}(x) \nu(x),
\end{equation}
with
\begin{equation}
\begin{split}
     \nu(x)=&P_{\sf{act}} g(x) \left(\hat{G}_2(x)-\frac{1}{2}\hat{G}_1(x)g(x)\right)\nonumber\\
    &+\frac{G_{\sf{th}} ((B_0-B_1x)g^2(x)-2\hat{G}_2(x) )}{2(B_0-B_1x)^2}.
\end{split}
\end{equation}

Since $f(x)$ is concave for $x \in [S_{\min}, S_{\max}]$, the optimal number of selected features in problem \eqref{prob:surrogate}, denoted as $S^*$, can be determined by evaluating the nearest (feasible) integers below and above $x^*$. The value that maximizes the objective function $f(x)$ is then selected, provided that $\nu(S_{\min})\cdot \nu(S_{\max}) \leq 0$. Otherwise, if $  \nu(S_{\min})\cdot \nu(S_{\max} )\ge 0$, the optimal number of selected features is one of the endpoints, i.e., $S^*= \argmax_{S\in\{S_{\min},S_{\max}\}}f(x)$.
This completes the proof.

\bibliography{Ref}

@article{peng2024learning,
  title={Learning resource allocation policy: Vertex-GNN or edge-GNN?},
  author={Peng, Yao and Guo, Jia and Yang, Chenyang},
  journal={IEEE Trans. Mach. Learn. Commun. Netw.},
  volume={2},
  pages={190--209},
  year={2024},
  publisher={IEEE}
}

@article{lin2023pushing,
 author  = {Z. Lin and G. Qu and Q. Chen and X. Chen and Z. Chen and K. Huang},
  journal = {IEEE Commun. Mag.},
  title   = {Pushing Large Language Models to the 6G Edge: Vision, Challenges, and Opportunities},
  year    = {2025},
  volume  = {63},
  number  = {9},
  pages   = {52--59}
}

@ARTICLE{Yang2026BatchSizeFL,
  author  = {H. Yang and Z. Wang and K. Huang},
  journal = {IEEE Trans. Commun.},
  title   = {Optimal Batch-Size Control for Low-Latency Federated Learning With Device Heterogeneity},
  year    = {2026},
  volume  = {74},
  number  = {},
  pages   = {5232--5247},
  doi     = {10.1109/TCOMM.2026.3666674}
}

@ARTICLE{Chen_tcom2023,
	author={Chen, Qian and Meng, Weixiao and Han, Shuai and Li, Cheng and Quek, Tony Q. S.},
	journal={IEEE Trans. Commun.}, 
	title={Coverage Analysis of {SAGIN} With Sectorized Beam Pattern Under Shadowed-Rician Fading Channels}, 
	year={2023},
	month=aug,
	volume={71},
	number={8},
	pages={4988-5004},
	doi={10.1109/TCOMM.2023.3280219}}

@article{chen2024space,
	  author  = {Q. Chen and Z. Wang and X. Chen and J. Wen and D. Zhou and S. Ji and M. Sheng and K. Huang},
  journal = {Engineering},
  title   = {Space--Ground Fluid {AI} for 6{G} Edge Intelligence},
  year    = {2025},
  volume  = {54},
  number  = {},
  pages   = {14--19},
  doi     = {10.1016/j.eng.2025.06.009}
}

@inproceedings{robot_arm,
  title={Using simulation and domain adaptation to improve efficiency of deep robotic grasping},
  author={Bousmalis, Konstantinos and Irpan, Alex and Wohlhart, Paul and Bai, Yunfei and Kelcey, Matthew and Kalakrishnan, Mrinal and Downs, Laura and Ibarz, Julian and Pastor, Peter and Konolige, Kurt and others},
  booktitle={IEEE Int. Conf. Rob. Autom. (ICRA)},
  pages={4243--4250},
  year={2018},
  address={Brisbane, Australia}
}

@article{GX-CM-2020,
  title={Toward an intelligent edge: Wireless communication meets machine learning},
  author={Zhu, Guangxu and Liu, Dongzhu and Du, Yuqing and You, Changsheng and Zhang, Jun and Huang, Kaibin},
  journal={IEEE Commun. Mag.},
  volume={58},
  number={1},
  pages={19--25},
  year={2020},
  publisher={IEEE}
}

@article{zhiyan_ISEA_survey,
 author  = {Liu, Zhiyan and Chen, Xu and Wu, Hai and Wang, Zhanwei and Chen, Xianhao and Niyato, Dusit and Huang, Kaibin},
  journal = {IEEE Commun. Surveys Tuts.},
  title   = {Integrated Sensing and Edge AI: Realizing Intelligent Perception in {{6G}}},
  year    = {2026},
  volume  = {28},
  pages   = {2725--2770},
  doi     = {10.1109/COMST.2025.3592989}
}

@inproceedings{biswas2016intelligent,
  author={S. P. Biswas and P. Roy and N. Patra and A. Mukherjee and N. Dey},
  title={Intelligent Traffic Monitoring System},
  booktitle={Proc. Int. Conf. Comput. Commun. Technol. (IC3T)},
  volume={2},
  pages={535--545},
  year={2016},
  address={Allahabad, India},
}

@article{Liva-TCOM-2019,
   title={Short packets over block-memoryless fading channels: Pilot-assisted or noncoherent transmission?},
  author={{\"O}stman, Johan and Durisi, Giuseppe and Str{\"o}m, Erik G and Co{\c{s}}kun, Mustafa C and Liva, Gianluigi},
  journal={IEEE Trans. Commun.},
  volume={67},
  number={2},
  pages={1521--1536},
  year={2018},
  publisher={IEEE}
}

@article{Petar-TCOM-2017,
   title={Downlink transmission of short packets: framing and control information revisited},
  author={Trillingsgaard, Kasper Fl{\o}e and Popovski, Petar},
  journal={IEEE Trans. Commun.},
  volume={65},
  number={5},
  pages={2048--2061},
  year={2017},
  publisher={IEEE}
}

@article{Quek-TCOM-2018,
   title={Joint uplink and downlink resource configuration for ultra-reliable and low-latency communications},
  author={She, Changyang and Yang, Chenyang and Quek, Tony QS},
  journal={IEEE Trans. Commun.},
  volume={66},
  number={5},
  pages={2266--2280},
  year={2018},
  publisher={IEEE}
}

@article{Schmeink-JSAC-2018,
   title={{SWIPT}-enabled relaying in {IoT} networks operating with finite blocklength codes},
  author={Hu, Yulin and Zhu, Yao and Gursoy, M Cenk and Schmeink, Anke},
  journal={IEEE J. Sel. Areas Commun.},
  volume={37},
  number={1},
  pages={74--88},
  year={2018},
  publisher={IEEE}
}

@article{Nallanathan-TWC-2020,
  title={Joint power and blocklength optimization for {URLLC} in a factory automation scenario},
  author={Ren, Hong and Pan, Cunhua and Deng, Yansha and Elkashlan, Maged and Nallanathan, Arumugam},
  journal={IEEE Trans. Wireless Commun.},
  volume={19},
  number={3},
  pages={1786--1801},
  year={2019},
  publisher={IEEE}
}

@InProceedings{simonyan2015deep,
  author       = {Karen Simonyan and Andrew Zisserman},
  title        = {Very Deep Convolutional Networks for Large-Scale Image Recognition},
  booktitle    = {Proc. Int. Conf. Learn. Represent. (ICLR)},
  year         = {2015},
  month = {May~7-9 },
  address = {San Diego, CA, USA}
}

@InProceedings{ModelNet-Ref,
author = {Su, Hang and Maji, Subhransu and Kalogerakis, Evangelos and Learned-Miller, Erik},
title = {Multi-View Convolutional Neural Networks for {3D} Shape Recognition},
booktitle = {Proc. IEEE Int. Conf. Comput. Vision (ICCV)},
month = {Dec. 7-13},
year = {2015},
address={Santiago, Chile},
}

@article{ZW2024ultra-LoLa,
author  = {Z. Wang and A. E. Kal{\o}r and Y. Zhou and P. Popovski and K. Huang},
  journal = {IEEE Trans. Wireless Commun.},
  title   = {Ultra-Low-Latency Edge Inference for Distributed Sensing},
  year    = {2026},
  volume  = {25},
  number  = {},
  pages   = {1908--1922},
  doi     = {10.1109/TWC.2025.3593802}
}

@article{ZJ-CoM-2020,
  title={Communication-computation trade-off in resource-constrained edge inference},
  author={Shao, Jiawei and Zhang, Jun},
  journal={IEEE Commun. Mag.},
  volume={58},
  number={12},
  pages={20--26},
  year={2020},
  publisher={IEEE}
}

@ARTICLE{Task-oriented_NextG_2023,
  author={Sagduyu, Yalin E. and Ulukus, Sennur and Yener, Aylin},
  journal={IEEE Wireless Commun.}, 
  title={Task-Oriented Communications for {NextG}: End-to-end Deep Learning and {AI} Security Aspects}, 
  year={2023},
  volume={30},
  number={3},
  pages={52-60}
}

@ARTICLE{Task-oriented_SC_2023,
  author={Ma, Shuai and Qiao, Weining and Wu, Youlong and Li, Hang and Shi, Guangming and Gao, Dahua and Shi, Yuanming and Li, Shiyin and Al-Dhahir, Naofal},
  journal={IEEE Trans. Wireless Commun.}, 
  title={Task-Oriented Explainable Semantic Communications}, 
  year={2023},
  volume={22},
  number={12},
  pages={9248-9262}
}

@ARTICLE{Task-based_ADC_2021,
  author={Neuhaus, Peter and Shlezinger, Nir and Dörpinghaus, Meik and Eldar, Yonina C. and Fettweis, Gerhard},
  journal={IEEE Trans. Signal Process.}, 
  title={Task-Based Analog-to-Digital Converters}, 
  year={2021},
  volume={69},
  pages={5403-5418},
}

@ARTICLE{Task_oriented_6G_2023,
  author={Shi, Yuanming and Zhou, Yong and Wen, Dingzhu and Wu, Youlong and Jiang, Chunxiao and Letaief, Khaled B.},
  journal={IEEE Wireless Commun.}, 
  title={Task-Oriented Communications for 6{G}: Vision, Principles, and Technologies}, 
  year={2023},
  volume={30},
  number={3},
  pages={78-85}
}

@INPROCEEDINGS{who2com,
  author={Liu, Yen-Cheng and Tian, Junjiao and Ma, Chih-Yao and Glaser, Nathan and Kuo, Chia-Wen and Kira, Zsolt},
  booktitle={Proc. IEEE Int. Conf. Robot. Automat. (ICRA)}, 
  title={Who2com: Collaborative Perception via Learnable Handshake Communication}, 
  year={2020},
  pages={6876-6883},
 }

@article{Zhiyan-JASC-2023,
  title={Resource allocation for multiuser edge inference with batching and early exiting},
  author={Liu, Zhiyan and Lan, Qiao and Huang, Kaibin},
  journal={IEEE J. Sel. Areas Commun.},
  volume={41},
  number={4},
  pages={1186--1200},
  year={2023},
  publisher={IEEE}
}

@article{Zhiyan-AirPooling,
  author={Liu, Zhiyan and Lan, Qiao and Kalør, Anders E. and Popovski, Petar and Huang, Kaibin},
  journal={IEEE Trans. Wireless Commun.}, 
  title={Over-the-Air Multi-View Pooling for Distributed Sensing}, 
  year={2024},
  volume={23},
  number={7},
  pages={7652-7667},
}

@inproceedings{Yang-CVPR-2014,
  title={Real-time simultaneous pose and shape estimation for articulated objects using a single depth camera},
  author={Ye, Mao and Yang, Ruigang},
  booktitle={Proc. IEEE Conf. Comput. Vis. Pattern Recognit. (CVPR)},
    address={Columbus, OH, USA},
    month={Jun. 23--28},
  year={2014}
}

@inproceedings{figueroa2019semi,
   title={Semi-supervised learning using deep generative models and auxiliary tasks},
  author={Figueroa, Jhosimar Arias},
  booktitle={Proc. Adv. Neural Inf. Process. Syst. (NeurIPS) Workshop},
  address={Vancouver, Canada},
  month={Dec. 13--14},
  year={2019}
}

@ARTICLE{zhao2023joint,
  author={Zhao, Yizhe and Hu, Jie and Yang, Kun and Wei, Xiaohui},
  journal={IEEE Trans. Veh. Technol.}, 
  title={A Joint Communication and Control System for {URLLC} in Industrial {IoT}}, 
  year={2023},
  volume={72},
  number={11},
  pages={15074-15079},
}

@article{wen2023taskOTA,
  title={Task-oriented over-the-air computation for multi-device edge {AI}},
  author={Wen, Dingzhu and Jiao, Xiang and Liu, Peixi and Zhu, Guangxu and Shi, Yuanming and Huang, Kaibin},
  journal={IEEE Trans. Wireless Commun.},
  year={2023},
  volume={23},
  number={3},
  pages={2039--2053},
  publisher={IEEE}
}

@article{wen2023task,
  title={Task-oriented sensing, computation, and communication integration for multi-device edge {AI}},
  author={Wen, Dingzhu and others},
  journal={IEEE Trans. Wireless Commun.},
  year={2023},
  publisher={IEEE},
  volume={23},
  number={3},
  pages={2486--2502},
}

@article{RN291,
  author={Lan, Qiao and Zeng, Qunsong and Popovski, Petar and Gündüz, Deniz and Huang, Kaibin},
  journal={IEEE Trans. Wireless Commun.}, 
  title={Progressive Feature Transmission for Split Classification at the Wireless Edge}, 
  year={2023},
  volume={22},
  number={6},
  pages={3837-3852},
}

@inproceedings{carlini2017towards,
  title={Towards evaluating the robustness of neural networks},
  author={Carlini, Nicholas and Wagner, David},
  booktitle={IEEE Symp. Secur. Privacy (SP)},
  pages={39--57},
  year={2017},
  address={San Jose, CA, USA},
}

@inproceedings{RN366,
   author = {Yan, Zheyu and Qin, Yifan and Wen, Wujie and Hu, Xiaobo Sharon and Shi, Yiyu},
   title = {Improving Realistic Worst-Case Performance of {NVCiM DNN} Accelerators Through Training with Right-Censored Gaussian Noise},
   booktitle = {Proc. IEEE/ACM Int. Conf. Comput.-Aided Des. (ICCAD)},
   year={2023},
   address={San Francisco, CA, USA},
}

@article{Andrews2011,
  title={A tractable approach to coverage and rate in cellular networks},
  author={Andrews, Jeffrey G and Baccelli, Fran{\c{c}}ois and Ganti, Radha Krishna},
  journal={IEEE Trans. Commun.},
  volume={59},
  number={11},
  pages={3122--3134},
  year={2011},
  month={Nov.},
  publisher={IEEE}
}

@article{Haenggi2009,
  author    = {Haenggi, M. and Andrews, J. G. and Baccelli, F. and Dousse, O. and Franceschetti, M.},
  title     = {Stochastic geometry and random graphs for the analysis and design of wireless networks},
  journal   = {IEEE J. Sel. Areas Commun.},
  volume    = {27},
  number    = {7},
  pages     = {1029--1046},
  month     = {Sep.},
  year      = {2009},
}

@article{QSFL,
   author = {Zeng, Qunsong and Du, Yuqing and Huang, Kaibin and Leung, Kin K.},
   title = {Energy-Efficient Resource Management for Federated Edge Learning With {CPU-GPU} Heterogeneous Computing},
   journal = {IEEE Trans. Wireless Commun.},
   volume = {20},
   number = {12},
   pages = {7947-7962},
   ISSN = {1536-1276},
   year = {2021},
}

@article{Dhillon2012,
  author    = {Dhillon, H. S. and Ganti, R. K. and Baccelli, F. and Andrews, J. G.},
  title     = {Modeling and Analysis of {K-Tier} Downlink Heterogeneous Cellular Networks},
  journal   = {IEEE J. Sel. Areas Commun.},
  volume    = {30},
  number    = {3},
  pages     = {550--560},
  month     = {April},
  year      = {2012},
}

@book{feller1991introduction,
    author    = {W. Feller},
  title     = {An Introduction to Probability Theory and Its Applications, vol. 2},
  publisher = {John Wiley \& Sons},
  year      = {1991}
}

@article{abdi2010principal,
  title={Principal component analysis},
  author={Abdi, Herv{\'e} and Williams, Lynne J},
  journal={Wiley Interdiscip. Rev.: Comput. Stat.},
  volume={2},
  number={4},
  pages={433--459},
  year={2010},
  publisher={Wiley Online Library}
}

@ARTICLE{ZW_spectrum,
  author={Wang, Zhanwei and Huang, Kaibin and Eldar, Yonina C.},
  journal={IEEE Trans. Wireless Commun.}, 
  title={Spectrum Breathing: Protecting Over-the-Air Federated Learning Against Interference}, 
  year={2024},
  volume={23},
  number={8},
  pages={10058-10071},
}

@article{zhang2019robustness,
  title={Robustness of Neural Networks: A Probabilistic and Practical Approach},
  author={Zhang, X. and others},
  journal={IEEE Trans. Neural Netw. Learn. Syst.},
  volume={30},
  number={9},
  pages={2540--2554},
  year={2019},
  publisher={IEEE}
}

@article{zeng2024knowledge,
  title={Knowledge-based ultra-low-latency semantic communications for robotic edge intelligence},
  author={Zeng, Qunsong and Wang, Zhanwei and Zhou, You and Wu, Hai and Yang, Lin and Huang, Kaibin},
  journal={IEEE Trans. Commun.},
  year={2024},
  publisher={IEEE}
}

@inproceedings{feature_pruning,
 author    = {Y. Zhu and C. Li and B. Luo and J. Tang and X. Wang},
  title     = {Dense Feature Aggregation and Pruning for {RGBT} Tracking},
  booktitle = {Proc. 27th {ACM} Int. Conf. Multimedia ({ACM} MM)},
  pages     = {465--472},
  address   = {Nice, France},
  year      = {2019},
  doi       = {10.1145/3343031.3350928}
}

@article{goldsmith1997capacity,
  title={Capacity of fading channels with channel side information},
  author={Goldsmith, Andrea J and Varaiya, Pravin P},
  journal={IEEE Trans. Inf. Theory},
  volume={43},
  number={6},
  pages={1986--1992},
  year={1997},
  publisher={IEEE}
}

@article{goldsmith1997variable,
  title={Variable-rate variable-power {MQAM} for fading channels},
  author={Goldsmith, Andrea J and Chua, Soon-Ghee},
  journal={IEEE Trans. Commun.},
  volume={45},
  number={10},
  pages={1218--1230},
  year={1997},
  publisher={IEEE}
}

@article{goldsmith1998adaptive,
  title={Adaptive coded modulation for fading channels},
  author={Goldsmith, Andrea J and Chua, Soon-Ghee},
  journal={IEEE Trans. Commun.},
  volume={46},
  number={5},
  pages={595--602},
  year={1998},
  publisher={IEEE}
}

@article{vishwanath2003adaptive,
  title={Adaptive turbo-coded modulation for flat-fading channels},
  author={Vishwanath, Sriram and Goldsmith, Andrea},
  journal={IEEE Trans. Commun.},
  volume={51},
  number={6},
  pages={964--972},
  year={2003},
  publisher={IEEE}
}

@book{goldsmith2005wireless,
  title={Wireless communications},
  author={Goldsmith, Andrea},
  year={2005},
  publisher={Cambridge university press}
}

@book{Alouni_book,
  title={Digital communication over fading channels},
  author={Simon, Marvin K and Alouini, Mohamed-Slim},
  year={2004},
  publisher={John Wiley \& Sons}
}

@article{alouini1999capacity,
  title={Capacity of Rayleigh fading channels under different adaptive transmission and diversity-combining techniques},
  author={Alouini, M-S and Goldsmith, Andrea J},
  journal={IEEE Trans. Veh. Technol.},
  volume={48},
  number={4},
  pages={1165--1181},
  year={1999},
  publisher={IEEE}
}

@article{hasna2003outage,
  title={Outage probability of multihop transmission over Nakagami fading channels},
  author={Hasna, Mazen Omar and Alouini, M-S},
  journal={IEEE Commun. Lett.},
  volume={7},
  number={5},
  pages={216--218},
  year={2003},
  publisher={IEEE}
}

@article{alouini2000adaptive,
  title={Adaptive modulation over Nakagami fading channels},
  author={Alouini, Mohamed-Slim and Goldsmith, Andrea J},
  journal={Wireless Pers. Commun.},
  volume={13},
  pages={119--143},
  year={2000},
  publisher={Springer}
}

@article{zheng2003diversity,
  title={Diversity and multiplexing: A fundamental tradeoff in multiple-antenna channels},
  author={Zheng, Lizhong and Tse, David N. C.},
  journal={IEEE Trans. Inf. Theory},
  volume={49},
  number={5},
  pages={1073--1096},
  year={2003},
  publisher={IEEE}
}

@book{tse2005fundamentals,
  title={Fundamentals of wireless communication},
  author={Tse, David and Viswanath, Pramod},
  year={2005},
  publisher={Cambridge university press}
}

@article{li_zeng_zhou_chen_2020,
  title={Edge {AI}: On-demand accelerating deep neural network inference via edge computing},
  author={Li, En and Zeng, Liekang and Zhou, Zhi and Chen, Xu},
  journal={IEEE Trans. Wireless Commun.},
  volume={19},
  number={1},
  pages={447--457},
  year={2019},
  publisher={IEEE}
}

@article{tang1999effect,
  title={Effect of channel estimation error on {M-QAM} {BER} performance in Rayleigh fading},
  author={Tang, Xiaoyi and Alouini, M-S and Goldsmith, Andrea J},
  journal={IEEE Trans. Commun.},
  volume={47},
  number={12},
  pages={1856--1864},
  year={1999},
  publisher={IEEE}
}

@misc{NVIDIA_Jetson_TX2,
  author       = {{NVIDIA Corp.}},
  title        = {{NVIDIA Jetson TX2 Series Module Datasheet}},
  howpublished = {[Online]. Available: \url{https://www.nvidia.com/en-us/autonomous-machines/embedded-systems/jetson-tx2/}},
}

@article{Mao2017MECsurvey,
  author  = {Yuyi Mao and Changsheng You and Jun Zhang and Kaibin Huang and Khaled B. Letaief},
  title   = {A Survey on Mobile Edge Computing: The Communication Perspective},
  journal = {IEEE Commun. Surveys Tuts.},
  volume  = {19},
  number  = {4},
  pages   = {2322--2358},
  year    = {2017},
  doi     = {10.1109/COMST.2017.2745201}
}

@article{You2017MECO,
  author  = {Changsheng You and Kaibin Huang and Haejoon Chae and Byounghoon Kim},
  title   = {Energy-Efficient Resource Allocation for Mobile-Edge Computation Offloading},
  journal = {IEEE Trans. Wireless Commun.},
  volume  = {16},
  number  = {3},
  pages   = {1397--1411},
  year    = {2017},
  month   = mar,
  doi     = {10.1109/TWC.2016.2633522}
}

@MISC{Wang2025AirBreathSensing,
  author       = {Z. Wang and M. Cui and H. Yang and Q. Zeng and M. Sheng and K. Huang},
  title        = {AirBreath Sensing: Protecting Over-the-Air Distributed Sensing Against Interference},
  howpublished = {arXiv:2508.11267},
  year         = {2025},
}
\bibliographystyle{IEEEtran}

\newpage

\begin{IEEEbiography}[{\includegraphics[width=1in,height=1.25in,clip,keepaspectratio]{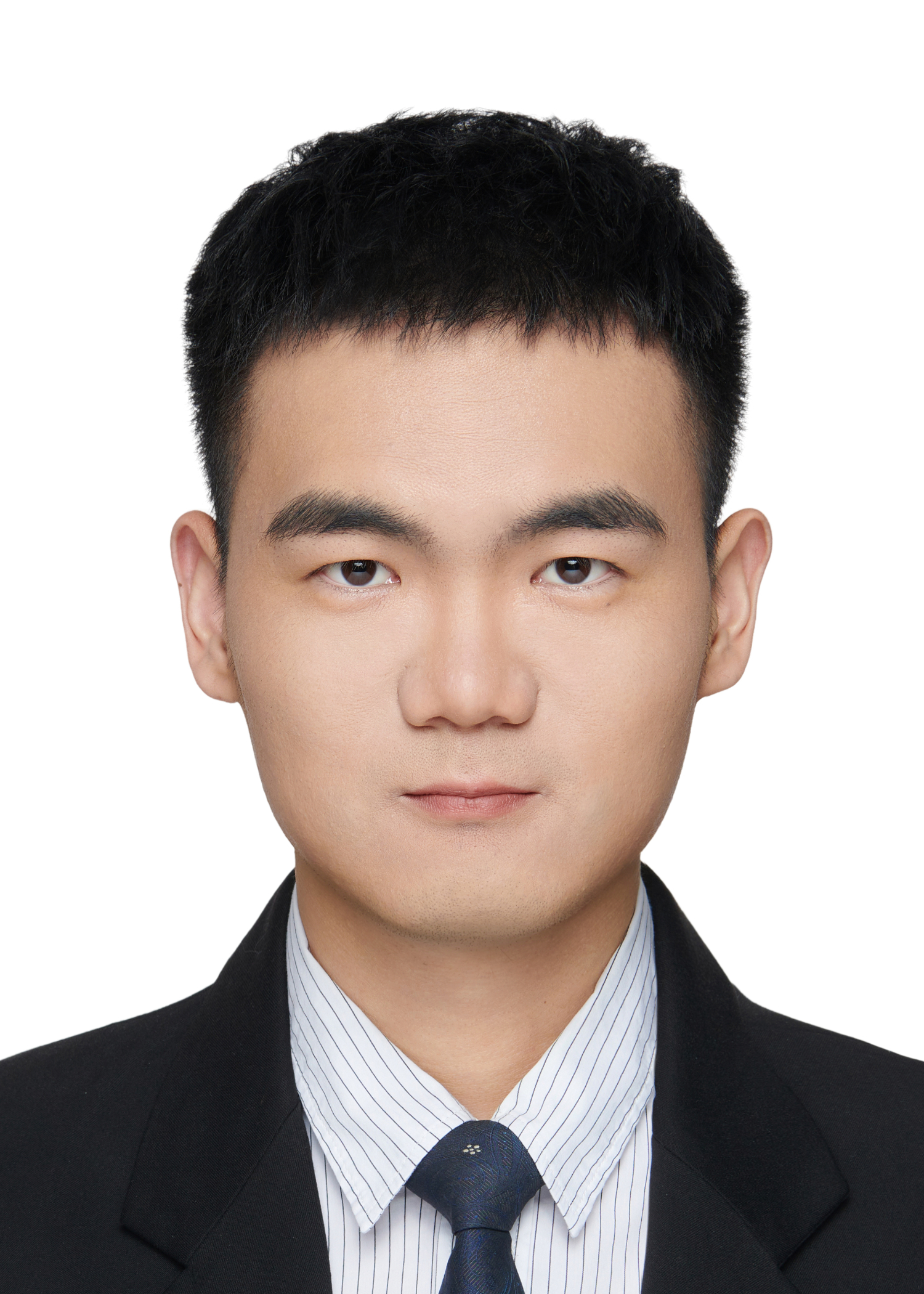}}]{Zhanwei Wang} (Member, IEEE) received the B.Eng. degree in Information Engineering and the M.Eng. degree in Information and Communication Engineering from Xidian University, Xi'an, China, in 2018 and 2021, respectively. He received the Ph.D. degree in Electrical and Computer Engineering from The University of Hong Kong (HKU), Hong Kong SAR, China. He is currently a Research Assistant Professor with the Department of Electrical and Electronic Engineering, HKU. His research interests include wireless communications, edge intelligence, atomic receivers, and space artificial intelligence (Space AI).
\end{IEEEbiography}

\begin{IEEEbiography}[{\includegraphics[width=1in,height=1.25in, clip,keepaspectratio]{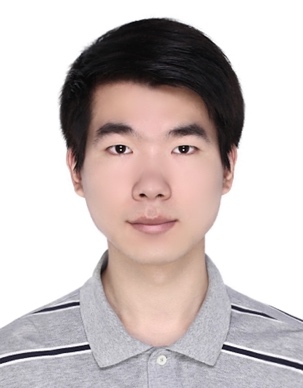}}]{Qunsong Zeng} (Member, IEEE) received the B.S. degree in physics (Yan Jici Talent Students Program) from the University of Science and Technology of China (USTC) and the Ph.D. degree in electrical and electronic engineering from The University of Hong Kong (HKU). He is currently a Research Assistant Professor with the Department of Electrical and Computer Engineering, The University of Hong Kong, Hong Kong. He was a recipient of HKU Foundation Award, P.K. Yu Memorial Scholarship, Y S and Christabel Lung Scholarship, etc. His research interests include edge intelligence, in-memory baseband processing and quantum receivers. He served as the co-chair for IEEE PIMRC’26 (WS-12), IEEE PIMRC’25 (WS-10) and IEEE/CIC ICCC’25 (WS-22).
\end{IEEEbiography}

\begin{IEEEbiography}[{\includegraphics[width=1in,height=1.25in,clip,keepaspectratio]{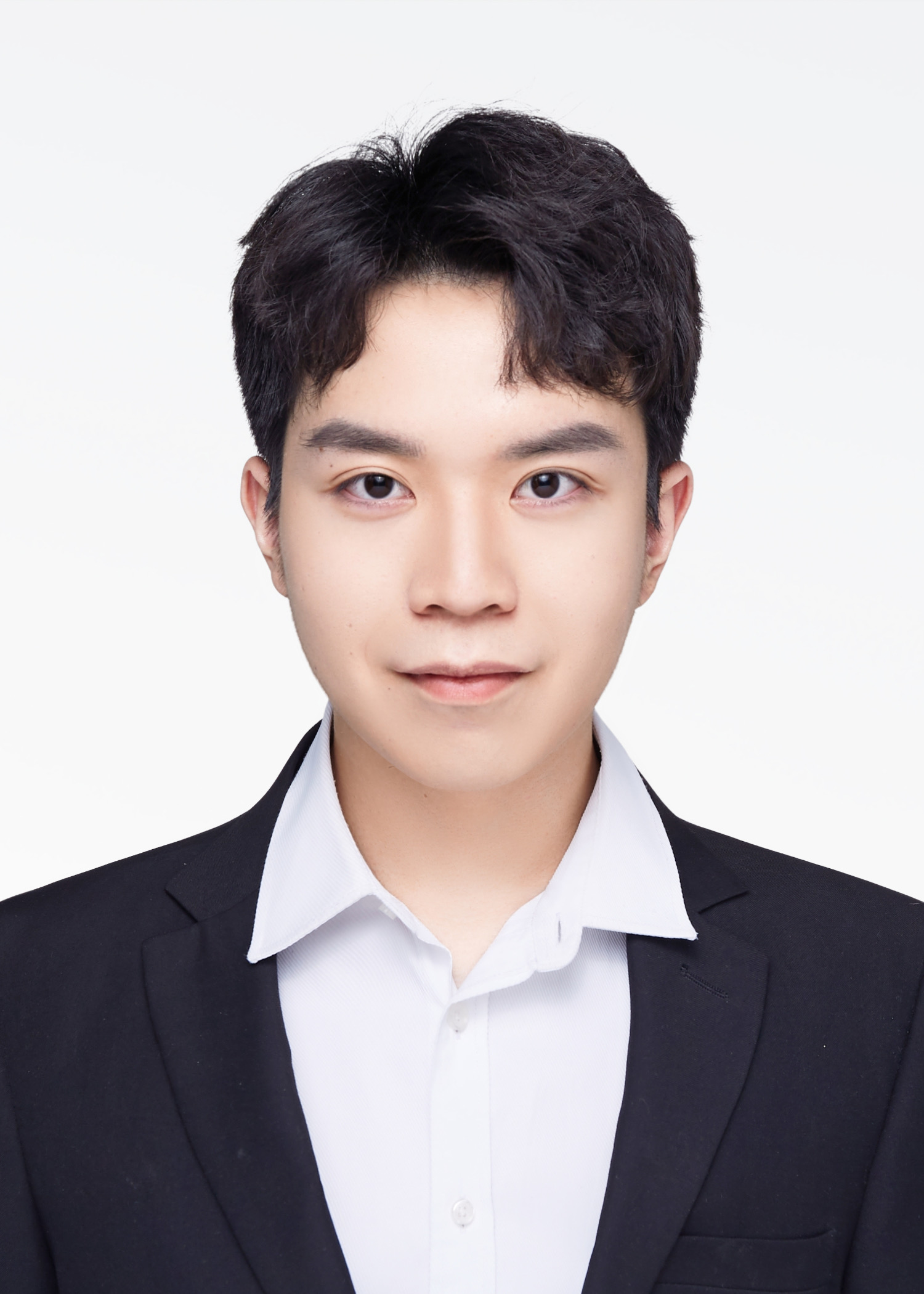}}]{Haotian Zheng} (Graduate Student Member, IEEE) received the B.Eng. degree in Electronic Information Engineering from Xidian University, Xi'an, China, in 2024. He is currently pursuing the Ph.D. degree with the Department of Electrical and Computer Engineering, The University of Hong Kong (HKU), Hong Kong SAR, China. His research interests include edge intelligence and space artificial intelligence (Space AI).
\end{IEEEbiography}

\begin{IEEEbiography}[{\includegraphics[width=1in,height=1.25in,clip,keepaspectratio]
{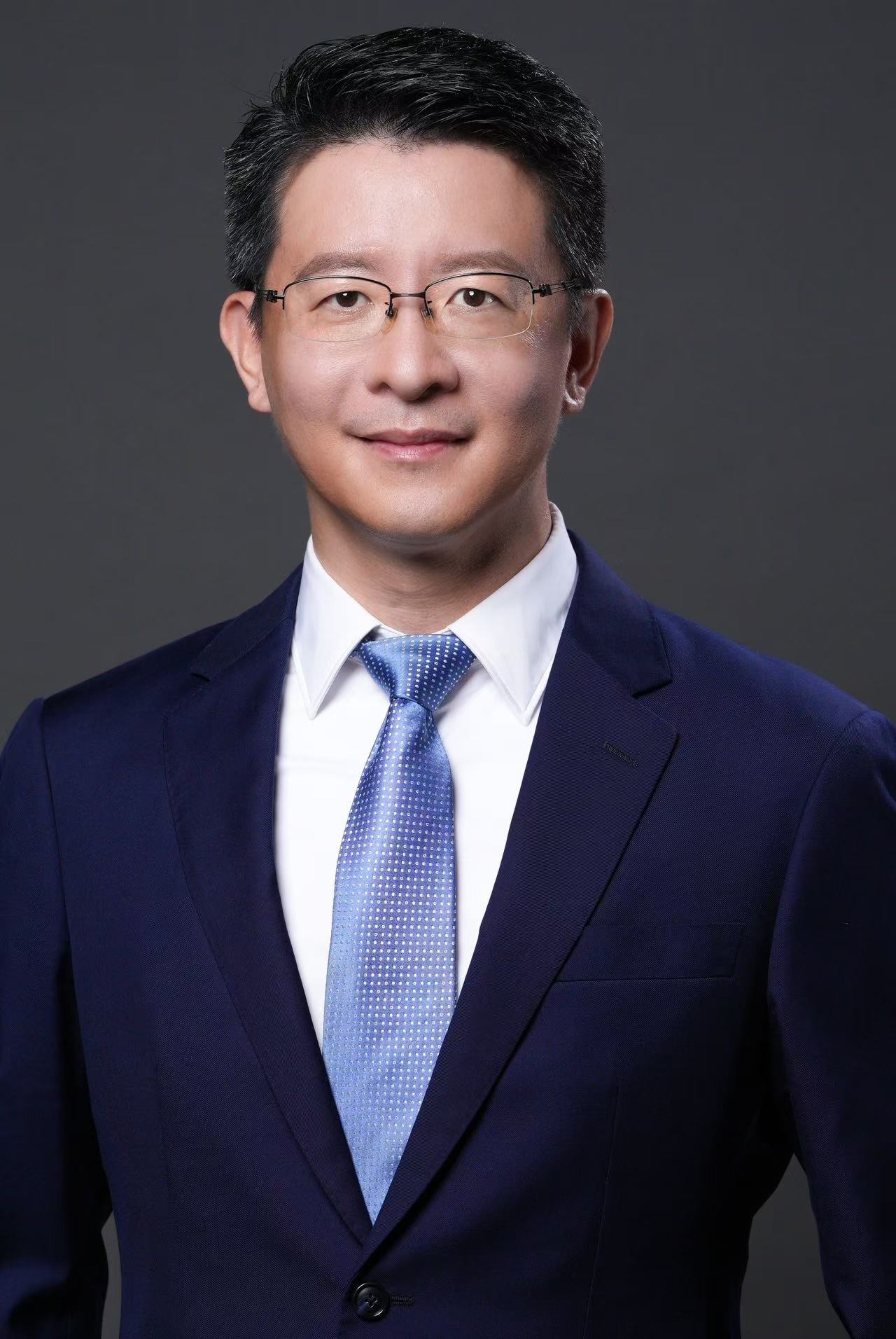}}]{Kaibin Huang} (Fellow, IEEE)  received the B.Eng. and M.Eng. degrees from the National University of Singapore and the Ph.D. degree from The University of Texas at Austin, all in electrical engineering. He is the Philip K H Wong Wilson K L Wong Professor in Electrical Engineering and the Department Head at the Dept. of Electrical and Computer Engineering, The University of Hong Kong (HKU), Hong Kong. His work was recognized with seven Best Paper awards from the IEEE Communication Society. He has served on the editorial boards of five major journals in the area of wireless communications and co-edited twelve journal special issues. He has been named as a Highly Cited Researcher by Clarivate in the last seven years (2019-2025) and an AI 2000 Most Influential Scholar (Top 30 in Internet of Things) in 2023-2025. He received the 2025 IEEE Wireless Communications Technical Committee Recognition Award. He is a member of the Engineering Panel of Hong Kong Research Grants Council.  He was an IEEE Distinguished Lecturer (2020-2022). He is a Croucher Senior Research Fellow (2026), Fellow of the IEEE (2021), and Fellow of the U.S. National Academy of Inventors (2024). 
\end{IEEEbiography}

\end{document}